%% file: paper.tex
\documentclass[preprint]{sig-alternate}

\usepackage{url}
\usepackage{subfigure}
\usepackage{xspace}
\usepackage{graphicx}
\usepackage[T1]{fontenc}
\usepackage{rustic}
\usepackage{color}
\usepackage{tabularx}
\usepackage{breqn}
\usepackage{listings}
\usepackage{courier}
\usepackage{comment}
\usepackage{times}

\newcommand{\subparagraph}{}
\usepackage[small,compact]{titlesec}

\newcommand{\ie}{{\em i.e.,}~}
\newcommand{\eg}{{\em e.g.,}~}
\newcommand{\system}{Shark\xspace}
\newcommand{\pde}{partial DAG execution\xspace}
\newcommand{\Pde}{Partial DAG execution\xspace}
\newcommand{\PDE}{Partial DAG Execution\xspace}
\newcommand{\PDEac}{PDE\xspace}

\newenvironment{packed_enum}{
\begin{enumerate}
  \setlength{\itemsep}{4pt}
  \setlength{\parskip}{0pt}
  \setlength{\parsep}{0pt}
}{\end{enumerate}}

\usepackage{etoolbox}
\makeatletter
\patchcmd{\maketitle}{\@copyrightspace}{}{}{}
\makeatother

\begin{document}

\title{\system: SQL and Rich Analytics at Scale}

\numberofauthors{6}
\author{
Reynold Xin,
Josh Rosen,
Matei Zaharia,\\
Michael J. Franklin,
Scott Shenker,
Ion Stoica\\\\
AMPLab, EECS, UC Berkeley\\
\affaddr{\{rxin, joshrosen, matei, franklin, shenker, istoica\}@cs.berkeley.edu}
}

\maketitle

\begin{abstract}
\system is a new data analysis system that marries query processing with complex analytics on
large clusters.
It leverages a novel distributed memory abstraction to provide a unified engine that can run SQL queries
and sophisticated analytics functions (\eg iterative machine learning) at scale, and efficiently recovers from failures
mid-query.
This allows \system to run SQL queries up to $100\times$ faster than Apache Hive, and machine learning
programs up to $100\times$ faster than Hadoop.
Unlike previous systems, \system shows that it is possible to achieve these speedups while retaining a
MapReduce-like execution engine, and the fine-grained fault tolerance properties that such engines provide.
It extends such an engine in several ways, including column-oriented in-memory storage and
dynamic mid-query replanning, to effectively execute SQL.
The result is a system that matches the speedups reported for MPP analytic databases over MapReduce,
while offering fault tolerance properties and complex analytics capabilities that they lack.
\end{abstract}




\input{intro}

\input{architecture}

\section{Engine Extensions}

In this section, we describe our modifications to the Spark engine to enable efficient execution of SQL queries.

\input{optimizations}

\input{memstore}


\input{machinelearning}

\input{implementation}

\input{experiments}

\input{discussion}

\input{related_work}

\input{conclusion}


\section{Acknowledgments}
We thank Cliff Engle, Harvey Feng, Shivaram Venkataraman, Ram Sriharsha,
Denny Britz, Antonio Lupher, Patrick Wendell, and Paul Ruan for their work on \system.
This research is supported in part by NSF CISE Expeditions award CCF-1139158, gifts from
Amazon Web Services, Google, SAP, Blue Goji, Cisco, Cloudera, Ericsson, General Electric,
Hewlett Packard, Huawei, Intel, Microsoft, NetApp, Oracle, Quanta, Splunk, VMware and
by DARPA (contract \#FA8650-11-C-7136).

\balancecolumns

\begin{small}
\bibliographystyle{abbrv}
\bibliography{paper}
\end{small}

\end{document}

%% file: intro.tex

\section{Introduction}
\label{sec:intro}

Modern data analysis faces a confluence of growing challenges.
First, data volumes are expanding dramatically, creating the need to scale out across clusters
of hundreds of commodity machines.
Second, this new scale increases the incidence of faults and stragglers (slow tasks),
complicating parallel database design.
Third, the \emph{complexity} of data analysis has also grown: modern data analysis employs
sophisticated statistical methods, such as machine learning algorithms, that go well beyond the roll-up and drill-down capabilities of traditional
enterprise data warehouse systems.
Finally, despite these increases in scale and complexity, users still expect to be able to query data at
interactive speeds.

To tackle the ``big data'' problem, two major lines of systems have recently been explored.
The first, composed of MapReduce~\cite{mapreduce} and various generalizations~\cite{dryad,tenzing}, offers
a fine-grained fault tolerance model suitable for large clusters,
where tasks on failed or slow nodes can be deterministically re-executed on other nodes. MapReduce is
also fairly general: it has been shown to be able to express many statistical and learning
algorithms~\cite{mapreduce-ml}.  
It also easily supports unstructured data and ``schema-on-read.''
However, MapReduce engines lack many of the features that make databases efficient, and have high
latencies of tens of seconds to hours.
Even systems that have significantly optimized MapReduce for SQL queries, such as Google's Tenzing~\cite{tenzing},
or that combine it with a traditional database on each node, such as HadoopDB~\cite{hadoopdb}, report a
minimum latency of 10 seconds.
As such, MapReduce approaches have largely been dismissed for interactive-speed queries~\cite{pavlo2009comparison},
and even Google is developing new engines for such workloads~\cite{dremel}.

Instead, most MPP analytic databases (\eg Vertica, Greenplum, Teradata) and several of the new
low-latency engines proposed for MapReduce environments (\eg Google Dremel~\cite{dremel}, Cloudera Impala~\cite{impala})
employ a coarser-grained recovery model, where an entire query has to be resubmitted if a
machine fails.\footnote{
Dremel provides fault tolerance within a query, but Dremel is limited
to aggregation trees instead of the more complex communication patterns in joins.
}
This works well for short queries where a retry is inexpensive, but faces significant challenges in
long queries as clusters scale up~\cite{hadoopdb}.
In addition, these systems often lack the rich analytics functions that are easy to implement in
MapReduce, such as machine learning and graph algorithms.
Furthermore, while it may be possible to implement some of these functions using UDFs, these algorithms
are often expensive, furthering the need for fault and straggler recovery for long queries.
Thus, most organizations tend to use other systems alongside MPP databases to perform complex analytics.

To provide an effective environment for big data analysis, we believe that processing systems will
need to support \emph{both} SQL and complex analytics efficiently, and to provide fine-grained fault
recovery across both types of operations.
This paper describes a new system that meets these goals, called \system.
\system is open source and compatible with Apache Hive, and has already been used at
web companies to speed up queries by 40--100$\times$.

\system builds on a recently-proposed distributed shared memory abstraction called
Resilient Distributed Datasets (RDDs)~\cite{spark} to perform most computations in memory while
offering fine-grained fault tolerance.
In-memory computing is increasingly important in large-scale analytics for two reasons.
First, many complex analytics functions, such as machine learning and graph algorithms,
are iterative, going over the data multiple times; thus, the fastest systems deployed
for these applications are in-memory~\cite{pregel,graphlab,spark}.
Second, even traditional SQL warehouse workloads exhibit strong temporal and spatial locality, because
more-recent fact table data and small dimension tables are read disproportionately often.
A study of Facebook's Hive warehouse and Microsoft's Bing analytics cluster showed that
over 95\% of queries in both systems could be served out of memory using just 64 GB/node
as a cache, even though each system manages more than 100 PB of total data~\cite{pacman}.

The main benefit of RDDs is an efficient mechanism for fault recovery.
Traditional main-memory databases support fine-grained updates to tables and replicate
writes across the network for fault tolerance, which is expensive on large commodity clusters.
In contrast, RDDs restrict the programming interface to \emph{coarse-grained} deterministic operators
that affect multiple data items at once, such as \emph{map}, \emph{group-by} and \emph{join},
and recover from failures by tracking the \emph{lineage} of each dataset and recomputing lost data.
This approach works well for data-parallel relational queries, and has also been shown to support
machine learning and graph computation~\cite{spark}.
Thus, when a node fails, \system can recover mid-query by rerunning the deterministic
operations used to build lost data partitions on other nodes, similar to MapReduce.
Indeed, it typically recovers within {seconds}, by parallelizing this work across the cluster.

\begin{figure}[t]
  \centering
  \includegraphics[width=\linewidth]{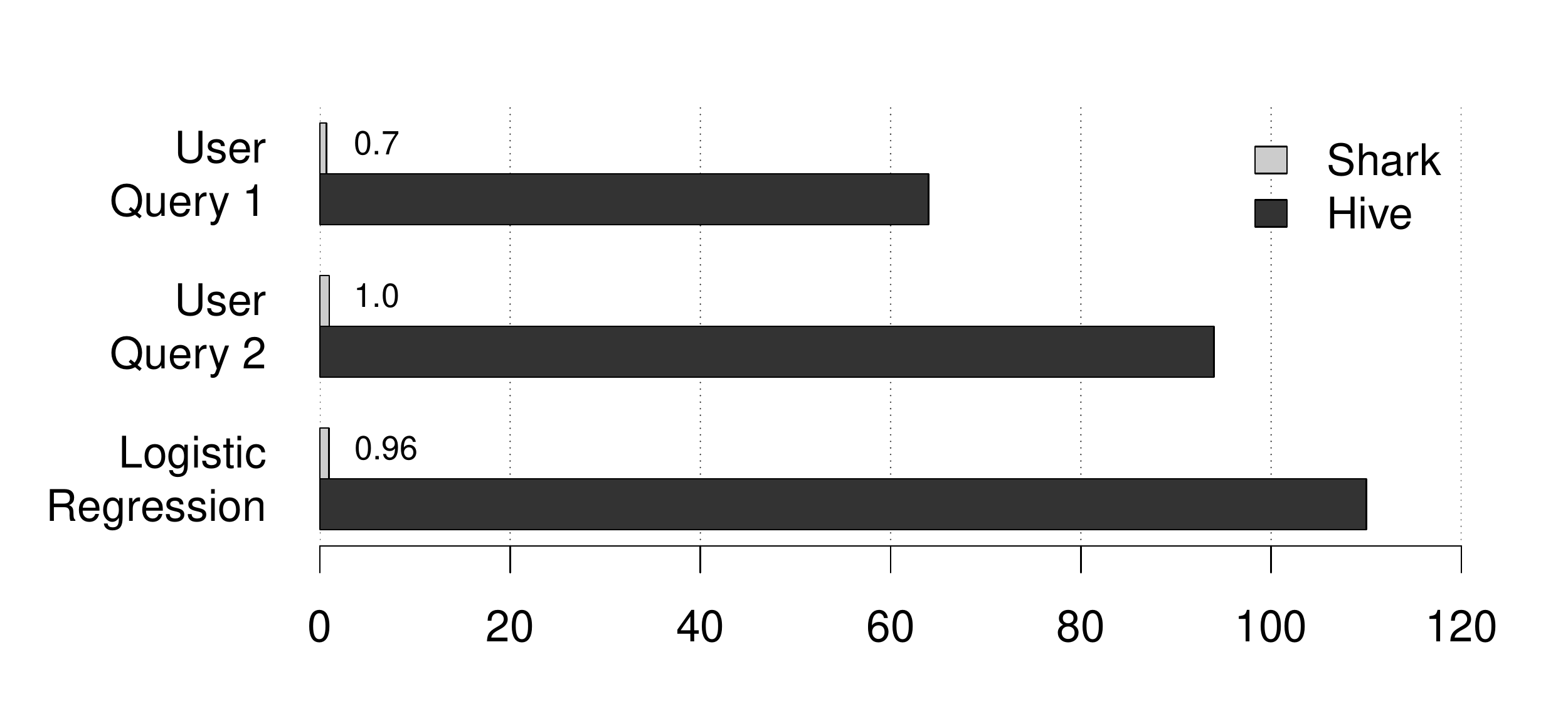}
  \vspace{-25pt}
  \caption{Performance of \system vs.~Hive/Hadoop on two SQL queries from an early user
  and one iteration of logistic regression (a classification algorithm that runs
  $\sim$10 such steps). Results measure the runtime (seconds) on a 100-node cluster.}
  \label{fig:top-chart}
\end{figure}

To run SQL efficiently, however, we also had to extend the RDD execution model, bringing
in several concepts from traditional analytical databases and some new ones. We started with
an existing implementation of RDDs called Spark~\cite{spark}, and added several features.
First, to store and process relational data efficiently, we implemented in-memory columnar
storage and columnar compression. This reduced both the data size and the processing time by
as much as $5\times$ over na\"ively storing the data in a Spark program in its original format.
Second, to optimize SQL queries based on the data characteristics even in the presence of
analytics functions and UDFs, we extended Spark with \emph{\PDE (\PDEac)}:
\system can reoptimize a running query after running the first few stages of its task DAG,
choosing better join strategies or the right degree of parallelism based on observed
statistics. Third, we leverage other properties of the Spark engine not present in traditional
MapReduce systems, such as control over data partitioning.

Our implementation of \system is compatible with Apache Hive \cite{hive}, supporting all
of Hive's SQL dialect and UDFs and allowing execution over unmodified
Hive data warehouses. It augments SQL with complex analytics
functions written in Spark, using Spark's Java, Scala or Python APIs. These functions can
be combined with SQL in a single execution plan, providing in-memory data sharing and fast
recovery across both types of processing.

Experiments show that using RDDs and the optimizations above, \system can answer SQL queries
up to $100\times$ faster than Hive, runs iterative machine learning algorithms
up to $100\times$ faster than Hadoop, and can recover from failures mid-query within seconds.
Figure~\ref{fig:top-chart} shows three sample results.
\system's speed is comparable to that of MPP databases in benchmarks like Pavlo et al.'s
comparison with MapReduce~\cite{pavlo2009comparison}, but it offers fine-grained recovery and
complex analytics features that these systems lack.

More fundamentally, our work shows that MapReduce-like execution models
can be applied effectively to SQL, and offer a promising way to
combine relational and complex analytics. In the course of presenting
of \system, we also explore why SQL engines over previous MapReduce runtimes, such as Hive, are slow,
and show how a combination of enhancements in \system (\eg \PDEac),
and engine properties that have not been optimized in MapReduce, such as the overhead of
launching tasks, eliminate many of the bottlenecks in traditional MapReduce systems.

%% file: architecture.tex

\section{System Overview}
\label{sec:arch}

\system is a data analysis system that supports both SQL query processing and machine learning functions.
We have chosen to implement \system to be compatible with Apache Hive.
It can be used to query an existing Hive warehouse and return results much faster, without modification to either the data or the queries.

Thanks to its Hive compatibility, \system can query data in any system that supports the Hadoop storage API, including HDFS and Amazon S3.
It also supports a wide range of data formats such as text, binary sequence files, JSON, and XML. It inherits
Hive's schema-on-read capability and nested data types~\cite{hive}.

In addition, users can choose to load high-value data into \system's memory store for fast analytics, as shown below:
\lstset{basicstyle=\small\ttfamily}
\begin{lstlisting}
CREATE TABLE latest_logs
  TBLPROPERTIES ("shark.cache"=true)
AS SELECT * FROM logs WHERE date > now()-3600;
\end{lstlisting}

\begin{figure}[t]
  \centering
  \includegraphics[width=\linewidth]{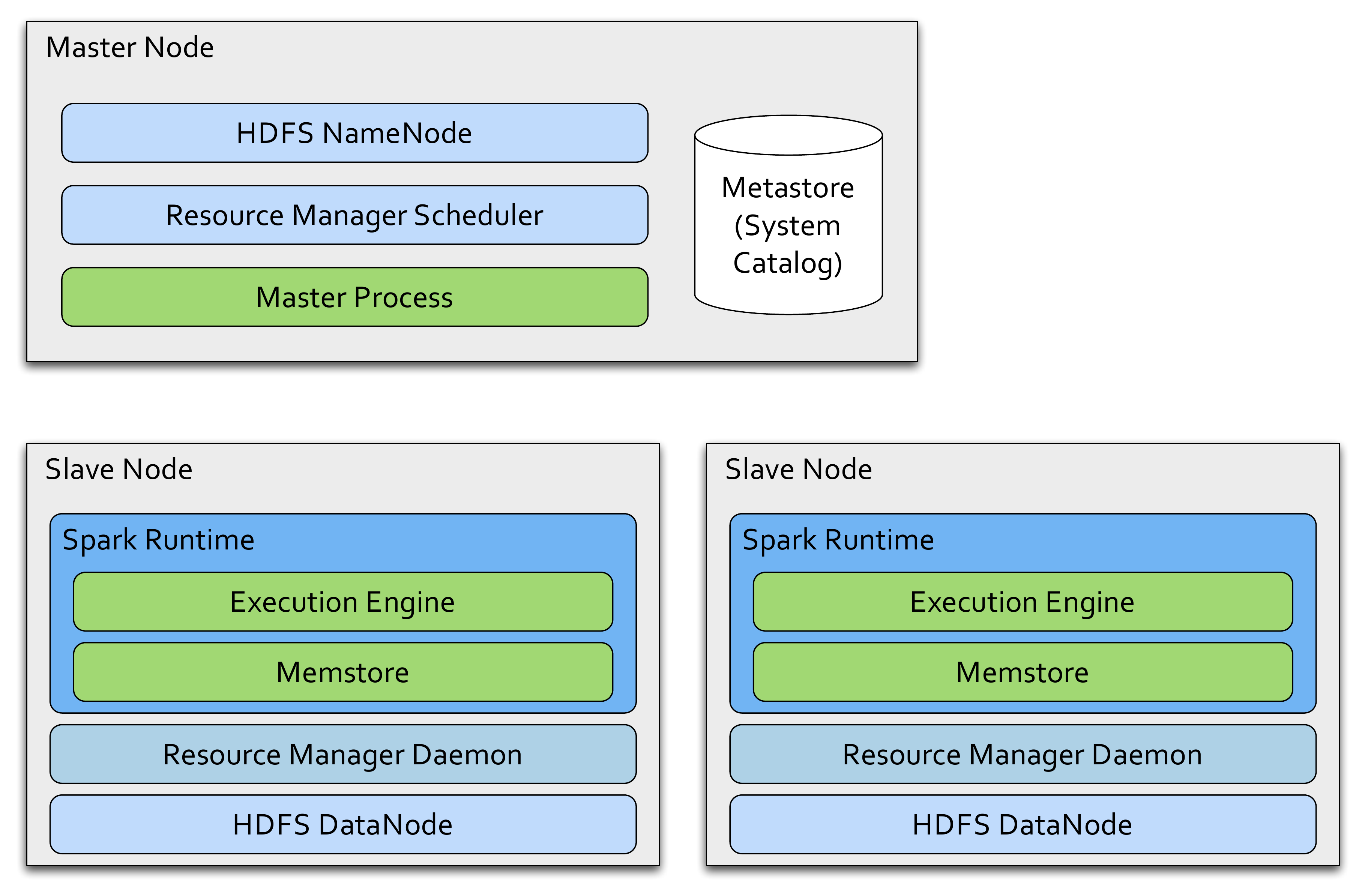}
  \caption{\system Architecture}
  \label{fig:arch}
\end{figure}

Figure \ref{fig:arch} shows the architecture of a \system cluster, consisting of a single master node and a number of slave nodes, with the warehouse metadata stored in an external transactional database.
It is built on top of Spark, a modern MapReduce-like cluster computing engine.
When a query is submitted to the master, \system compiles the query into operator tree represented as RDDs,
as we shall discuss in Section~\ref{sec:sql-over-rdds}.
These RDDs are then translated by Spark into a graph of tasks to execute on the slave nodes.

Cluster resources can optionally be allocated by a cluster resource manager (\eg Hadoop YARN or Apache Mesos) that provides resource sharing and isolation between different computing frameworks, allowing \system to coexist with engines like Hadoop.

In the remainder of this section, we cover the basics of Spark and the RDD programming model, followed by an explanation of how \system query plans are generated and run.

\subsection{Spark}

Spark is the MapReduce-like cluster computing engine used by \system.
Spark has several features that differentiate it from traditional MapReduce engines~\cite{spark}:

\begin{packed_enum}
  \item Like Dryad and Tenzing~\cite{dryad, tenzing}, it supports general computation DAGs, not just the two-stage MapReduce topology.

  \item It provides an in-memory storage abstraction called Resilient Distributed Datasets (RDDs)
  that lets applications keep data in memory across queries, and automatically reconstructs it
  after failures~\cite{spark}.

  \item The engine is optimized for low latency. It can efficiently
  manage tasks as short as 100 milliseconds on clusters of thousands of cores, while engines
  like Hadoop incur a latency of 5--10 seconds to launch each task.
\end{packed_enum}

RDDs are unique to Spark, and were essential to enabling
mid-query fault tolerance.
However, the other differences are important engineering elements that contribute to \system's performance.

On top of these features, we have also modified the Spark engine for \system to support
\emph{\pde,} that is, modification of the query plan DAG after only some of the stages have finished,
based on statistics collected from these stages. Similar to \cite{dewitt-mid-query}, we use this technique to optimize join algorithms and other
aspects of the execution mid-query, as we shall discuss in Section~\ref{sec:pde}.

\subsection{Resilient Distributed Datasets (RDDs)}

Spark's main abstraction is \emph{resilient distributed datasets} (RDDs), which are immutable,
partitioned collections that can be created through various data-parallel \emph{operators}
(\eg \emph{map}, \emph{group-by}, \emph{hash-join}).
Each RDD is either a collection stored in an external storage system, such as a file in HDFS,
or a derived dataset created by applying operators to other RDDs.
For example, given an RDD of (visitID, URL) pairs for visits to a website, we might
compute an RDD of (URL, count) pairs by applying a \emph{map} operator to turn each
event into an (URL, 1) pair, and then a \emph{reduce} to add the counts by URL.

In Spark's native API, RDD operations are invoked through a functional interface
similar to DryadLINQ~\cite{dryadlinq} in Scala, Java or Python.
For example, the Scala code for the query above is:

\lstset{basicstyle=\small\ttfamily}
\begin{lstlisting}
val visits = spark.hadoopFile("hdfs://...")
val counts = visits.map(v => (v.url, 1))
                   .reduceByKey((a, b) => a + b)
\end{lstlisting}

\begin{figure}[t]
  \centering
  \includegraphics[width=0.7\linewidth]{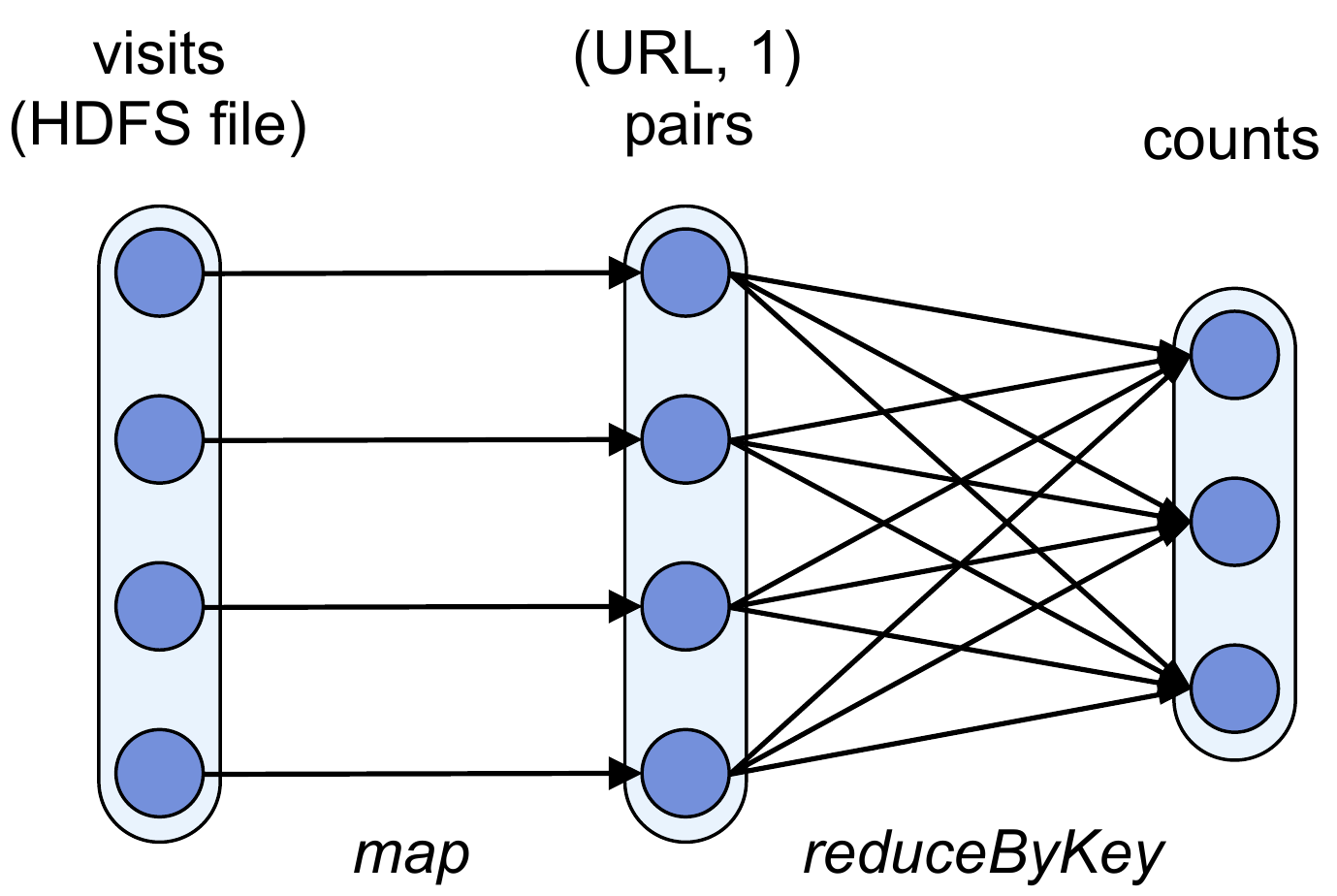}
  \caption{Lineage graph for the RDDs in our Spark example. Oblongs represent
  RDDs, while circles show partitions within a dataset. Lineage is tracked
  at the granularity of partitions.}
  \label{fig:rdd-example}
\end{figure}

RDDs can contain arbitrary data types as elements (since Spark runs on the JVM, these
elements are Java objects), and are automatically partitioned across the cluster, but they are
immutable once created, and they can only be created through Spark's deterministic parallel operators.
These two restrictions, however, enable highly efficient fault recovery.
In particular, instead of replicating each RDD across nodes for fault-tolerance, Spark remembers
the \emph{lineage} of the RDD (the graph of operators used to build it), and recovers lost partitions
by \emph{recomputing} them from base data~\cite{spark}.\footnote{
We assume that external files for RDDs representing external data do not change, or that we can
take a snapshot of a file when we create an RDD from it.
}
For example, Figure~\ref{fig:rdd-example} shows the lineage graph for the RDDs computed
above. If Spark loses one of the partitions in the (URL, 1) RDD, for example, it can recompute
it by rerunning the \emph{map} on just the corresponding partition of the input file.

The RDD model offers several key benefits our large-scale in-memory computing
setting. First, RDDs can be written at the speed of DRAM instead of the speed of the network, because
there is no need to replicate each byte written to another machine for fault-tolerance. DRAM in a modern server
is over $10\times$ faster than even a 10-Gigabit network. Second, Spark can keep just one copy
of each RDD partition in memory, saving precious memory over a replicated system, since it can always
recover lost data using lineage. Third, when a node fails, its lost RDD partitions can be rebuilt
\emph{in parallel} across the other nodes, allowing speedy recovery.\footnote{
To provide fault tolerance across ``shuffle'' operations like a parallel reduce, the execution engine also saves the
``map'' side of the shuffle in memory on the source nodes, spilling to disk if necessary.}
Fourth, even if a node is
just slow (a ``straggler''), we can recompute necessary partitions on other nodes because RDDs are
immutable so there are no consistency concerns with having two copies of a partition.
These benefits make RDDs attractive as the foundation for our relational processing in \system.

\subsection{Fault Tolerance Guarantees}

To summarize the benefits of RDDs explained above, \system provides the following fault tolerance
properties, which have been difficult to support in traditional MPP database designs:
\begin{packed_enum}
  \item \system can tolerate the loss of \emph{any set of worker nodes}. The execution engine will re-execute any lost tasks and recompute any lost RDD partitions using lineage.\footnote{
  Support for master recovery could also be added by reliabliy logging the RDD lineage graph and
  the submitted jobs, because this state is small, but we have not yet implemented this.
  }
  This is true even within a query: Spark will rerun any failed tasks, or lost
  dependencies of new tasks, without aborting the query.

  \item Recovery is parallelized across the cluster. If a failed node contained 100 RDD partitions, these can be rebuilt in
  parallel on 100 different nodes, quickly recovering the lost data.

  \item The deterministic nature of RDDs also enables straggler mitigation: if a task is slow, the system can launch a speculative ``backup copy'' of it on another node, as in MapReduce~\cite{mapreduce}.

  \item Recovery works even in queries that \emph{combine} SQL and machine learning UDFs
  (Section~\ref{sec:ml}), as these operations all compile into a single RDD lineage graph.

\end{packed_enum}

\subsection{Executing SQL over RDDs}
\label{sec:sql-over-rdds}

\system runs SQL queries over Spark using a three-step process similar to traditional RDBMSs: query parsing, logical plan generation, and physical plan generation.

Given a query, \system uses the Hive query compiler to parse the query and generate an abstract syntax tree.
The tree is then turned into a logical plan and basic logical optimization, such as predicate pushdown, is applied.
Up to this point, \system and Hive share an identical approach. Hive would then convert the operator into a physical plan consisting of multiple MapReduce stages.
In the case of \system, its optimizer applies additional rule-based optimizations, such as pushing \texttt{LIMIT} down to individual partitions, and creates a physical plan consisting of transformations on RDDs rather than MapReduce jobs.
We use a variety of operators already present in Spark, such as \emph{map} and \emph{reduce}, as well as new operators
we implemented for \system, such as broadcast joins.
Spark's master then executes this graph using standard MapReduce scheduling techniques, such placing tasks
close to their input data, rerunning lost tasks, and performing straggler mitigation~\cite{spark}.



While this basic approach makes it possible to run SQL over Spark, doing so \emph{efficiently} is challenging.
The prevalence of UDFs and complex analytic functions in \system's workload makes it difficult to determine an optimal
query plan at compile time, especially for new data that has not undergone ETL. In addition, even with such a plan,
na\"ively executing it over Spark (or other MapReduce runtimes) can be inefficient. In the next section, we discuss
several extensions we made to Spark to efficiently store relational data and run SQL, starting with a mechanism
that allows for \emph{dynamic}, statistics-driven re-optimization at run-time.


%% file: optimizations.tex
\subsection{Partial DAG Execution (\PDEac)}
\label{sec:pde}

Systems like \system and Hive are frequently used to query fresh data that has not undergone a data loading process.
This precludes the use of static query optimization techniques that rely on accurate a priori data statistics, such as statistics maintained by indices.
The lack of statistics for fresh data, combined with the prevalent use of UDFs, necessitates dynamic approaches to query optimization.

To support dynamic query optimization in a distributed setting, we extended Spark to support  \emph{\pde} (\PDEac), a technique that allows dynamic alteration of query plans based on data statistics collected at run-time.

We currently apply \pde at blocking ``shuffle" operator boundaries where data is exchanged and repartitioned, since these are typically the most expensive operations in \system.
By default, Spark materializes the output of each map task in memory before a shuffle, spilling it to disk as necessary.
Later, reduce tasks fetch this output.

PDE modifies this mechanism in two ways. First, it gathers customizable statistics at global and per-partition granularities while materializing map output.
Second, it allows the DAG to be altered based on these statistics, either by choosing different operators or altering their parameters (such as their degrees of parallelism).

These statistics are customizable using a simple, pluggable accumulator API. Some example statistics include:
\begin{packed_enum}
    \item Partition sizes and record counts, which can be used to detect skew.
    \item Lists of ``heavy hitters,'' \ie items that occur frequently in the dataset.
    \item Approximate histograms, which can be used to estimate partitions' data's distributions.
\end{packed_enum}

These statistics are sent by each worker to the master, where they are aggregated and presented to the optimizer.
For efficiency, we use lossy compression to record the statistics, limiting their size to 1--2 KB per task.
For instance, we encode partition sizes (in bytes) with logarithmic encoding, which can represent sizes of up to 32 GB using only one byte with at most 10\% error.
The master can then use these statistics to perform various run-time optimizations, as we shall discuss next.


\Pde complements existing adaptive query optimization techniques that typically run in a single-node system~\cite{eddies, dewitt-mid-query, query-scrambling}, as we can use existing techniques to dynamically optimize the local plan \emph{within} each node, and use \PDEac to optimize the global structure of the plan at stage boundaries.
This fine-grained statistics collection, and the optimizations that it enables, differentiates PDE from graph rewriting features in previous systems, such as DryadLINQ~\cite{dryadlinq}.





\subsubsection{Join Optimization}

\begin{figure}[t]
  \centering
  \includegraphics[width=\linewidth]{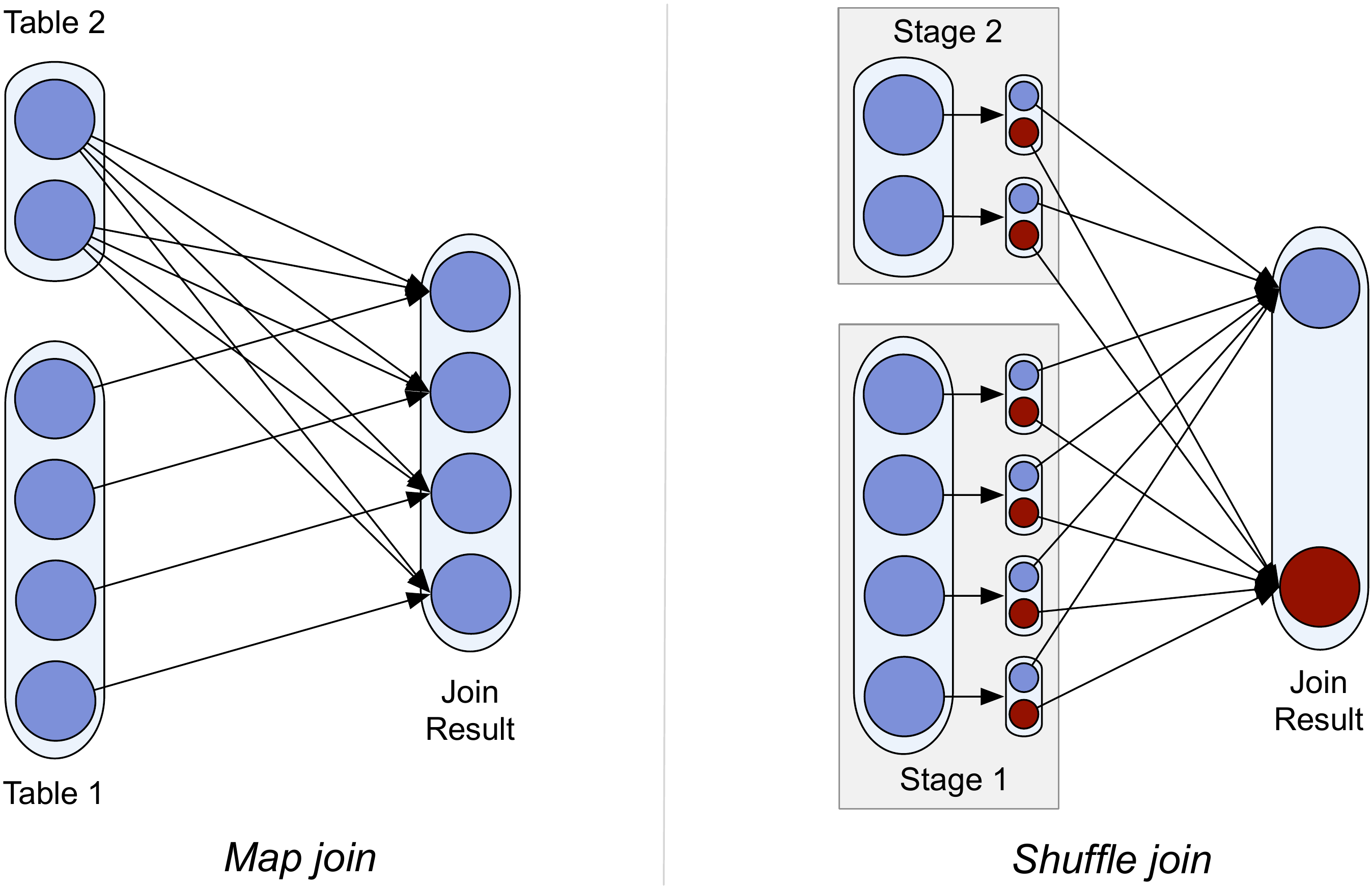}
  \vspace{-10pt}
  \caption{Data flows for map join and shuffle join.  Map join broadcasts the small table to all large table partitions, while shuffle join repartitions and shuffles both tables.}
  \label{fig:map-vs-shuffle-join}
\end{figure}

\Pde can be used to perform several run-time optimizations for join queries.

Figure~\ref{fig:map-vs-shuffle-join} illustrates two communication patterns for MapReduce-style joins.
In \emph{shuffle join}, both join tables are hash-partitioned by the join key.
Each reducer joins corresponding partitions using a local join algorithm, which is chosen by each reducer based on run-time statistics.
If one of a reducer's input partitions is small, then it constructs a hash table over the small partition and probes it using the large partition.
If both partitions are large, then a symmetric hash join is performed by constructing hash tables over both inputs.

In \emph{map join}, also known as \emph{broadcast join}, a small input table is broadcast to all nodes, where it is joined with each partition of a large table.
This approach can result in significant cost savings by avoiding an expensive repartitioning and shuffling phase.

Map join is only worthwhile if some join inputs are small, so \system uses \pde to select the join strategy at run-time based on its inputs' exact sizes.
By using sizes of the join inputs gathered at run-time, this approach works well even with input tables that have no prior statistics, such as intermediate results.


Run-time statistics also inform the join tasks' scheduling policies.
If the optimizer has a prior belief that a particular join input will be small, it will schedule that task before other join inputs and decide to perform a map-join if it observes that the task's output is small.
This allows the query engine to avoid performing the pre-shuffle partitioning of a large table once the optimizer has decided to perform a map-join.


\subsubsection{Skew-handling and Degree of Parallelism}
\Pde can also be used to determine operators' degrees of parallelism and to mitigate skew.

The degree of parallelism for reduce tasks can have a large performance impact: launching too few reducers may overload reducers' network connections and exhaust their memories, while launching too many may prolong the job due to task scheduling overhead.
Hive's performance is especially sensitive to the number of reduce tasks, due to  Hadoop's large scheduling overhead.

Using \pde, \system can use individual partitions' sizes to determine the number of reducers at run-time by coalescing many small, fine-grained partitions into fewer coarse partitions that are used by reduce tasks.
To mitigate skew, fine-grained partitions are assigned to coalesced partitions using a greedy bin-packing heuristic that attempts to equalize coalesced partitions' sizes \cite{closer}.
This offers performance benefits, especially when good bin-packings exist.

Somewhat surprisingly, we discovered that \system can obtain similar performance improvement by running a larger number of reduce tasks.  We attribute this to Spark's low scheduling overhead.

%% file: memstore.tex
\subsection{Columnar Memory Store}
\label{sec:memstore}

In-memory computation is essential to low-latency query answering, given that memory's throughput is orders of magnitude higher than that of disks.
Na\"ively using Spark's memory store, however, can lead to undesirable performance.
\system implements a columnar memory store on top of Spark's memory store.

In-memory data representation affects both space footprint and read throughput.
A na\"ive approach is to simply cache the on-disk data in its native format, performing on-demand deserialization in the query processor.
This deserialization becomes a major bottleneck: in our studies, we saw that modern commodity CPUs can deserialize at a rate of only 200MB per second per core.

The approach taken by Spark's default memory store is to store data partitions as collections of JVM objects.
This avoids deserialization, since the query processor can directly use these objects, but leads to significant storage space overheads.
Common JVM implementations add 12 to 16 bytes of overhead per object.
For example, storing 270 MB of TPC-H lineitem table as JVM objects uses approximately 971 MB of memory, while a serialized representation requires only 289 MB, nearly three times less space.
A more serious implication, however, is the effect on garbage collection (GC).
With a 200 B record size, a 32 GB heap can contain 160 million objects.
The JVM garbage collection time correlates linearly with the number of objects in the heap, so it could take minutes to perform a full GC on a large heap.
These unpredictable, expensive garbage collections cause large variability in workers' response times.

\system stores all columns of primitive types as JVM primitive arrays.
Complex data types supported by Hive, such as \texttt{map} and \texttt{array}, are serialized and concatenated into a single byte array.
Each column creates only one JVM object, leading to fast GCs and a compact data representation.
The space footprint of columnar data can be further reduced by cheap compression techniques at virtually no CPU cost.
Similar to more traditional database systems \cite{cstore}, \system implements CPU-efficient compression schemes such as dictionary encoding, run-length encoding, and bit packing.

Columnar data representation also leads to better cache behavior, especially for for analytical queries that frequently compute aggregations on certain columns.

\subsection{Distributed Data Loading}
\label{sec:dataloading}

In addition to query execution, \system also uses Spark's execution engine for distributed data loading.
During loading, a table is split into small partitions, each of which is loaded by a Spark task.
The loading tasks use the data schema to extract individual fields from rows, marshals a partition of data into its columnar representation, and stores those columns in memory.

Each data loading task tracks metadata to decide whether each column in a partition should be compressed.
For example, the loading task will compress a column using dictionary encoding if its number of distinct values is below a threshold.
This allows each task to choose the best compression scheme for each partition, rather than conforming to a global compression scheme that might not be optimal for local partitions.
These local decisions do not require coordination among data loading tasks, allowing the load phase to achieve a maximum degree of parallelism, at the small cost of requiring each partition to maintain its own compression metadata.
It is important to clarify that an RDD's lineage does not need to contain the compression scheme and metadata for each partition.
The compression scheme and metadata are simply byproducts of the RDD computation, and can be deterministically recomputed along with the in-memory data in the case of failures.

As a result, \system can load data into memory at the aggregated throughput of the CPUs processing incoming data.

Pavlo et al.\cite{pavlo2009comparison} showed that Hadoop was able to perform data loading at 5 to 10 times the throughput of MPP databases.
Tested using the same dataset used in \cite{pavlo2009comparison}, \system provides the same throughput as Hadoop in loading data into HDFS.
\system is 5 times faster than Hadoop when loading data into its memory store.

\subsection{Data Co-partitioning}
\label{sec:controlled-partitioning}

In some warehouse workloads, two tables are frequently joined together.
For example, the TPC-H benchmark frequently joins the lineitem and order tables.
A technique commonly used by MPP databases is to co-partition the two tables based on their join key in the data loading process.
In distributed file systems like HDFS, the storage system is schema-agnostic, which prevents data co-partitioning.
\system allows co-partitioning two tables on a common key for faster joins in subsequent queries.
This can be accomplished with the \texttt{DISTRIBUTE BY} clause:
\lstset{basicstyle=\small\ttfamily}
\begin{lstlisting}
CREATE TABLE l_mem TBLPROPERTIES ("shark.cache"=true)
AS SELECT * FROM lineitem DISTRIBUTE BY L_ORDERKEY;

CREATE TABLE o_mem TBLPROPERTIES (
  "shark.cache"=true, "copartition"="l_mem")
AS SELECT * FROM order DISTRIBUTE BY O_ORDERKEY;
\end{lstlisting}

When joining two co-partitioned tables, \system's optimizer constructs a DAG that avoids the expensive shuffle and instead uses map tasks to perform the join.

\subsection{Partition Statistics and Map Pruning}
\label{sec:mappruning}

Data tend to be stored in some logical clustering on one or more columns.
For example, entries in a website's traffic log data might be grouped by users' physical locations, because logs are first stored in data centers that have the best geographical proximity to users.
Within each data center, logs are append-only and are stored in roughly chronological order.
As a less obvious case, a news site's logs might contain \texttt{news\_id} and \texttt{timestamp} columns that have strongly correlated values.
For analytical queries, it is typical to apply filter predicates or aggregations over such columns.
For example, a daily warehouse report might describe how different visitor segments interact with the website; this type of query naturally applies a predicate on timestamps and performs aggregations that are grouped by geographical location.
This pattern is even more frequent for interactive data analysis, during which drill-down operations are frequently performed.

\emph{Map pruning} is the process of pruning data partitions based on their natural clustering columns.
Since \system's memory store splits data into small partitions, each block contains only one or few logical groups on such columns, and \system can avoid scanning certain blocks of data if their values fall out of the query's filter range.

To take advantage of these natural clusterings of columns, \system's memory store on each worker piggybacks the data loading process to collect statistics.
The information collected for each partition include the range of each column and the distinct values if the number of distinct values is small (\ie enum columns).
The collected statistics are sent back to the master program and kept in memory for pruning partitions during query execution.

When a query is issued, \system evaluates the query's predicates against all partition statistics; partitions that do not satisfy the predicate are pruned and \system does not launch tasks to scan them.


We collected a sample of queries from the Hive warehouse of a video analytics company, and out of the 3833 queries we obtained, at least 3277 of them contain predicates that \system can use for map pruning.
Section~\ref{sec:exp} provides more details on this workload.

%% file: machinelearning.tex
\section{Machine Learning Support}
\label{sec:ml}

A key design goal of \system is to provide a single system capable of efficient SQL query processing and sophisticated machine learning.
Following the principle of pushing computation to data, \system supports machine learning as a first-class citizen.
This is enabled by the design decision to choose Spark as the execution engine and RDD as the main data structure for operators.
In this section, we explain \system's language and execution engine integration for SQL and machine learning.

Other research projects \cite{madlib, unified-db-analytics} have demonstrated that it is possible to express certain machine learning algorithms in SQL and avoid moving data out of the database.
The implementation of those projects, however, involves a combination of SQL, UDFs, and driver programs written in other languages.
The systems become obscure and difficult to maintain; in addition, they may sacrifice performance by performing expensive parallel numerical computations on traditional database engines that were not designed for such workloads.
Contrast this with the approach taken by \system, which offers in-database analytics that push computation to data, but does so using a runtime that is optimized for such workloads and a programming model that is designed to express machine learning algorithms.

\subsection{Language Integration}
In addition to executing a SQL query and returning its results, \system also allows queries to return the RDD representing the query plan. Callers to \system can then invoke distributed computation over the query result using the returned RDD.

As an example of this integration, Listing~\ref{listing:logReg} illustrates a data analysis pipeline that performs logistic regression over a user database.
Logistic regression, a common classification algorithm, searches for a hyperplane \emph{w} that best separates two sets of points (e.g. spammers and non-spammers).
The algorithm applies gradient descent optimization by starting with a randomized \emph{w} vector and iteratively updating it by moving along gradients towards an optimum value.

The program begins by using \texttt{sql2rdd} to issue a SQL query to retreive user information as a \texttt{TableRDD}.
It then performs feature extraction on the query rows and runs logistic regression over the extracted feature matrix.
Each iteration of \texttt{logRegress} applies a function of \emph{w} to all data points to produce a set of gradients, which are summed to produce a net gradient that is used to update \emph{w}.

\begin{figure}[t]
\lstset{basicstyle=\small\ttfamily, keywords={map, reduce, mapRows, sql2rdd}, frame=tb, caption=Logistic Regression Example, label=listing:logReg, captionpos=b}
\begin{lstlisting}
def logRegress(points: RDD[Point]): Vector {
  var w = Vector(D, _ => 2 * rand.nextDouble - 1)
  for (i <- 1 to ITERATIONS) {
    val gradient = points.map { p =>
      val denom = 1 + exp(-p.y * (w dot p.x))
      (1 / denom - 1) * p.y * p.x
    }.reduce(_ + _)
    w -= gradient
  }
  w
}

val users = sql2rdd("SELECT * FROM user u
   JOIN comment c ON c.uid=u.uid")

val features = users.mapRows { row =>
  new Vector(extractFeature1(row.getInt("age")),
             extractFeature2(row.getStr("country")),
             ...)}
val trainedVector = logRegress(features.cache())
\end{lstlisting}
\end{figure}

The highlighted \texttt{map}, \texttt{mapRows}, and \texttt{reduce} functions are automatically parallelized by \system to execute across a cluster, and the master program simply collects the output of the \texttt{reduce} function to update \emph{w}.

Note that this distributed logistic regression implementation in \system looks remarkably similar to a program implemented for a single node in the Scala language.
The user can conveniently mix the best parts of both SQL and MapReduce-style programming.

Currently, \system provides native support for Scala and Java, with support for Python in development.
We have modified the Scala shell to enable interactive execution of both SQL and distributed machine learning algorithms.
Because \system is built on top of the JVM, it is trivial to support other JVM languages, such as Clojure or JRuby.

We have implemented a number of basic machine learning algorithms, including linear regression, logistic regression, and k-means clustering.
In most cases, the user only needs to supply a \texttt{mapRows} function to perform feature extraction and can invoke the provided algorithms.

The above example demonstrates how machine learning computations can be performed on query results. Using RDD as the main data structure for query operators also enables the possibility of using SQL to query the results of machine learning computations in a single execution plan.

\subsection{Execution Engine Integration}

In addition to language integration, another key benefit of using RDDs as the data structure for operators is the execution engine integration.
This common abstraction allows machine learning computations and SQL queries to share workers and cached data without the overhead of data movement.

Because SQL query processing is implemented using RDDs, lineage is kept for the whole pipeline, which enables end-to-end fault tolerance for the entire workflow.
If failures occur during machine learning stage, partitions on faulty nodes will automatically be recomputed based on their lineage.

%% file: implementation.tex
\section{Implementation}
\label{sec:implementation}
\label{sec:memory-shuffle}

While implementing \system, we discovered that a number of engineering details had significant performance impacts.
Overall, to improve the query processing speed, one should minimize the tail latency of tasks and the CPU cost of processing each row.

\vspace{4pt}\noindent\textbf{Memory-based Shuffle:}
Both Spark and Hadoop write map output files to disk, hoping that they will remain in the OS buffer cache when reduce tasks fetch them.
In practice, we have found that the extra system calls and file system journaling adds significant overhead.
In addition, the inability to control when buffer caches are flushed leads to variability in the execution time of shuffle tasks.
A query's response time is determined by the last task to finish, and thus the increasing variability leads to long-tail latency, which significantly hurts shuffle performance.
We modified the shuffle phase to materialize map outputs in memory, with the option to spill them to disk.

\vspace{4pt}\noindent\textbf{Temporary Object Creation:}
It is easy to write a program that creates many temporary objects, which can burden the JVM's garbage collector.
For a parallel job, a slow GC at one task may slow the entire job.
\system operators and RDD transformations are written in a way that minimizes temporary object creations.

\vspace{4pt}\noindent\textbf{Bytecode Compilation of Expression Evaluators:}
In its current implementation, \system sends the expression evaluators generated by the Hive parser as part of the tasks to be executed on each row. By profiling \system, we discovered that for certain queries, when data is served out of the memory store the majority of the CPU cycles are wasted in interpreting these evaluators. We are working on a compiler to transform these expression evaluators into JVM bytecode, which can further increase the execution engine's throughput.

\vspace{4pt}\noindent\textbf{Specialized Data Structures:}
Using specialized data structures is another low-hanging optimization that we have yet to exploit.
For example, Java's hash table is built for generic objects. When the hash key is a primitive type, the use of specialized data structures can lead to more compact data representations, and thus better cache behavior.




%% file: experiments.tex

\section{Experiments}
\label{sec:exp}

We evaluated \system using four datasets:
\begin{packed_enum}
  \item Pavlo et al.~Benchmark: 2.1 TB of data reproducing Pavlo et al.'s comparison of MapReduce
  vs.~analytical DBMSs \cite{pavlo2009comparison}.
  \item TPC-H Dataset: 100 GB and 1 TB datasets generated by the DBGEN program~\cite{tpch}.
  \item Real Hive Warehouse: 1.7 TB of sampled Hive warehouse data from an early industrial user of \system.
  \item Machine Learning Dataset: 100 GB synthetic dataset to measure the performance of machine learning algorithms.
\end{packed_enum}

Overall, our results show that \system can perform up to $100\times$ faster than Hive, even though we have yet to implement some of the performance optimizations mentioned in the previous section.
In particular, \system provides comparable performance gains to those reported for MPP databases in Pavlo et al.'s
comparison \cite{pavlo2009comparison}.
In some cases where data fits in memory, \system exceeds the performance reported for MPP databases.

We emphasize that we are \emph{not} claiming that \system is fundamentally faster than MPP databases; there is no
reason why MPP engines could not implement the same processing optimizations as \system.
Indeed, our implementation has several disadvantages relative to commercial engines, such as running on the JVM.
Instead, we aim to show that it is possible to achieve comparable performance while retaining a MapReduce-like engine,
and the fine-grained fault recovery features that such engines provide.
In addition, \system can leverage this engine to perform high-speed machine learning functions on the same data,
which we believe will be essential in future analytics workloads.

\subsection{Methodology and Cluster Setup}

Unless otherwise specified, experiments were conducted on Amazon EC2 using 100 \texttt{m2.4xlarge} nodes.
Each node had 8 virtual cores, 68 GB of memory, and 1.6 TB of local storage.

The cluster was running 64-bit Linux 3.2.28, Apache Hadoop 0.20.205, and Apache Hive 0.9.
For Hadoop MapReduce, the number of map tasks and the number of reduce tasks per node were set to 8, matching the number of cores.
For Hive, we enabled JVM reuse between tasks and avoided merging small output files, which would take an extra step after each query to perform the merge.


We executed each query six times, discarded the first run, and report the average of the remaining five runs.
We discard the first run in order to allow the JVM's just-in-time compiler to optimize common code paths.
We believe that this more closely mirrors real-world deployments where the JVM will be reused by many queries.


\subsection{Pavlo et al. Benchmarks}

Pavlo et al.~compared Hadoop versus MPP databases and showed that Hadoop excelled at data ingress, but performed unfavorably in query execution~\cite{pavlo2009comparison}.
We reused the dataset and queries from their benchmarks to compare \system against Hive.

\begin{figure}[t]
  \centering
  \includegraphics[width=0.8\linewidth]{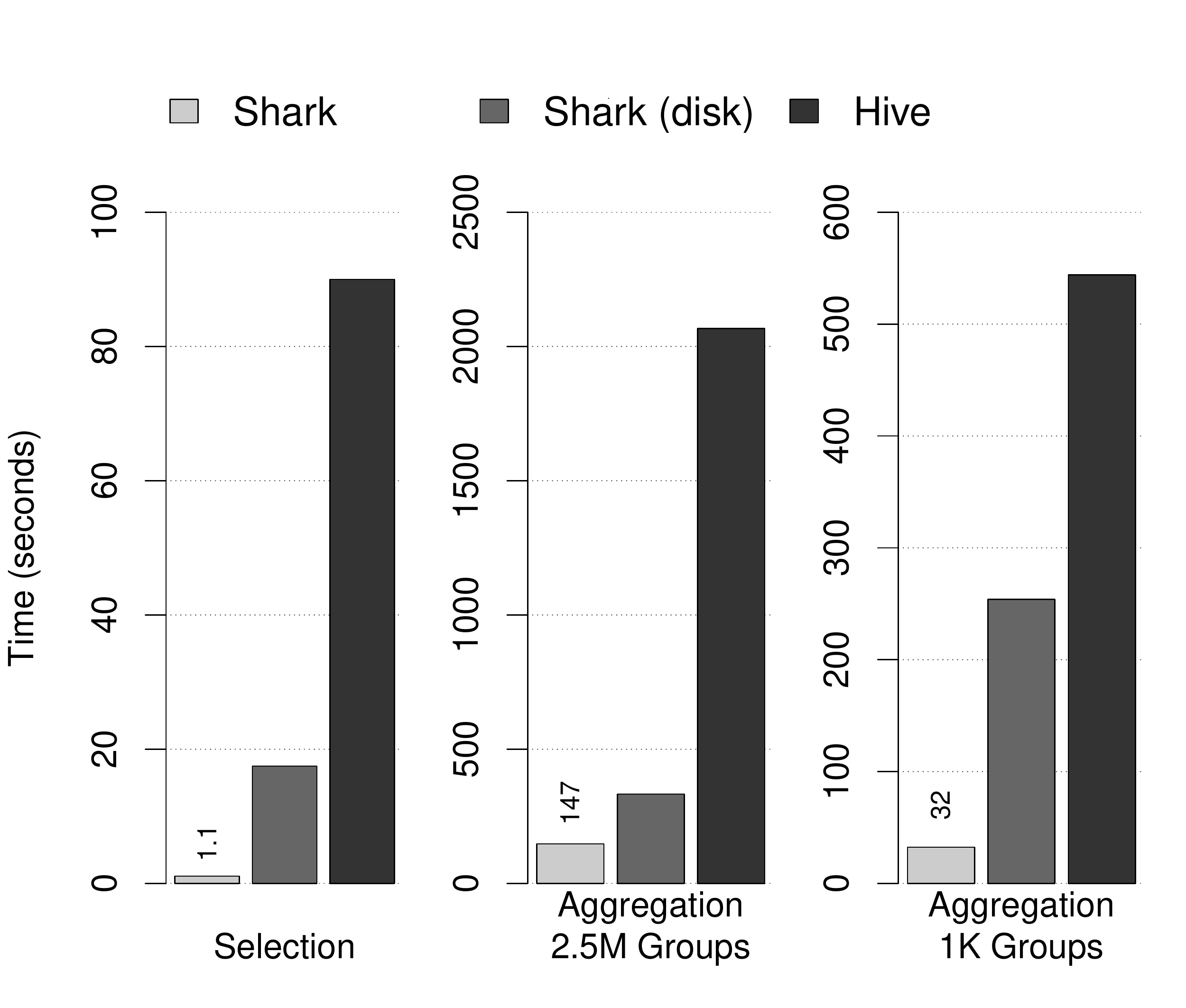}
  \vspace{-15pt}
  \caption{Selection and aggregation query runtimes (seconds) from Pavlo et al. benchmark}
  \label{fig:pavlo-groupby}
\end{figure}

The benchmark used two tables: a 1 GB/node rankings table, and a 20 GB/node uservisits table.
For our 100-node cluster, we recreated a 100 GB rankings table containing 1.8 billion rows and a 2 TB uservisits table containing 15.5 billion rows.
We ran the four queries in their experiments comparing \system with Hive and report the results in Figures \ref{fig:pavlo-groupby} and \ref{fig:pavlo-join}.
In this subsection, we hand-tuned Hive's number of reduce tasks to produce optimal results for Hive.
Despite this tuning, \system outperformed Hive in all cases by a wide margin.

\subsubsection{Selection Query}

The first query was a simple selection on the rankings table:
\lstset{basicstyle=\small\ttfamily}
\begin{lstlisting}
SELECT pageURL, pageRank
FROM rankings WHERE pageRank > X;
\end{lstlisting}

In \cite{pavlo2009comparison}, Vertica outperformed Hadoop by a factor of 10 because a clustered index was created for Vertica.
Even without a clustered index, \system was able to execute this query $80\times$ faster than Hive for in-memory data, and $5\times$ on data read from HDFS.

\subsubsection{Aggregation Queries}

The Pavlo et al. benchmark ran two aggregation queries:
\lstset{basicstyle=\small\ttfamily}
\begin{lstlisting}
SELECT sourceIP, SUM(adRevenue)
FROM uservisits GROUP BY sourceIP;

SELECT SUBSTR(sourceIP, 1, 7), SUM(adRevenue)
FROM uservisits GROUP BY SUBSTR(sourceIP, 1, 7);
\end{lstlisting}

In our dataset, the first query had two million groups and the second had approximately one thousand groups.
\system and Hive both applied task-local aggregations and shuffled the data to parallelize the final merge aggregation.
Again, \system outperformed Hive by a wide margin.
The benchmarked MPP databases perform local aggregations on each node, and then send all aggregates to a single query coordinator for the final merging; this performed very well when the number of groups was small, but performed worse with large number of groups.
The MPP databases' chosen plan is similar to choosing a single reduce task for \system and Hive.

\subsubsection{Join Query}

\begin{figure}[t]
  \centering
  \includegraphics[width=\linewidth]{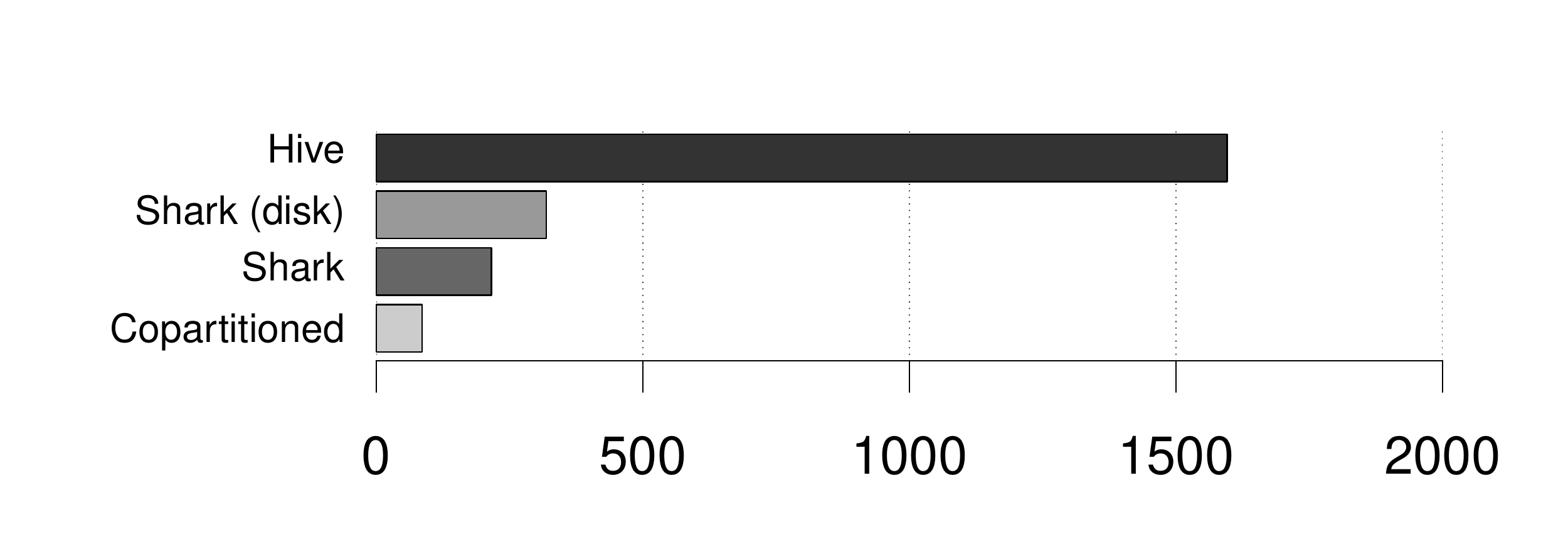}
  \vspace{-25pt}
  \caption{Join query runtime (seconds) from Pavlo benchmark}
  \label{fig:pavlo-join}
\end{figure}

The final query from Pavlo et al. involved joining the 2 TB uservisits table with the 100 GB rankings table.
\lstset{basicstyle=\small\ttfamily}
\begin{lstlisting}
SELECT INTO Temp sourceIP, AVG(pageRank),
SUM(adRevenue) as totalRevenue
FROM rankings AS R, uservisits AS UV
WHERE R.pageURL = UV.destURL
AND UV.visitDate BETWEEN Date('2000-01-15')
AND Date('2000-01-22')
GROUP BY UV.sourceIP;
\end{lstlisting}

Again, \system outperformed Hive in all cases.
Figure~\ref{fig:pavlo-join} shows that for this query, serving data out of memory did not provide much benefit over disk.
This is because the cost of the join step dominated the query processing.
Co-partitioning the two tables, however, provided significant benefits as it avoided shuffling data 2.1 TB of data during the join step.

\subsubsection{Data Loading}

Hadoop was shown by \cite{pavlo2009comparison} to excel at data loading, as its data loading throughput was five to ten times higher than that of MPP databases.
As explained in Section \ref{sec:arch}, \system can be used to query data in HDFS directly, which means its data ingress rate is at least as fast as Hadoop's.

After generating the 2 TB uservisits table, we measured the time to load it into HDFS and compared that with the time to load it into \system's memory store.
We found the rate of data ingress was $5\times$ higher in \system's memory store than that of HDFS.


\subsection{Micro-Benchmarks}

To understand the factors affecting \system's performance, we conducted a sequence of micro-benchmarks.
We generated 100 GB and 1 TB of data using the DBGEN program provided by TPC-H~\cite{tpch}.
We chose this dataset because it contains tables and columns of varying cardinality and can be used to create a myriad of micro-benchmarks for testing individual operators.

While performing experiments, we found that Hive and Hadoop MapReduce were very sensitive to the number of reducers set for a job.
Hive's optimizer automatically sets the number of reducers based on the estimated data size.
However, we found that Hive's optimizer frequently made the wrong decision, leading to incredibly long query execution times.
We hand-tuned the number of reducers for Hive based on characteristics of the queries and through trial and error.
We report Hive performance numbers for both optimizer-determined and hand-tuned numbers of reducers.
\system, on the other hand, was much less sensitive to the number of reducers and required minimal tuning.

\subsubsection{Aggregation Performance}

We tested the performance of aggregations by running group-by queries on the TPH-H lineitem table.
For the 100 GB dataset, lineitem table contained 600 million rows. For the 1 TB dataset, it contained 6 billion rows.

\begin{figure*}[ht]
  \centering
  \begin{minipage}[b]{0.45\linewidth}
    \centering
    \includegraphics[width=\linewidth]{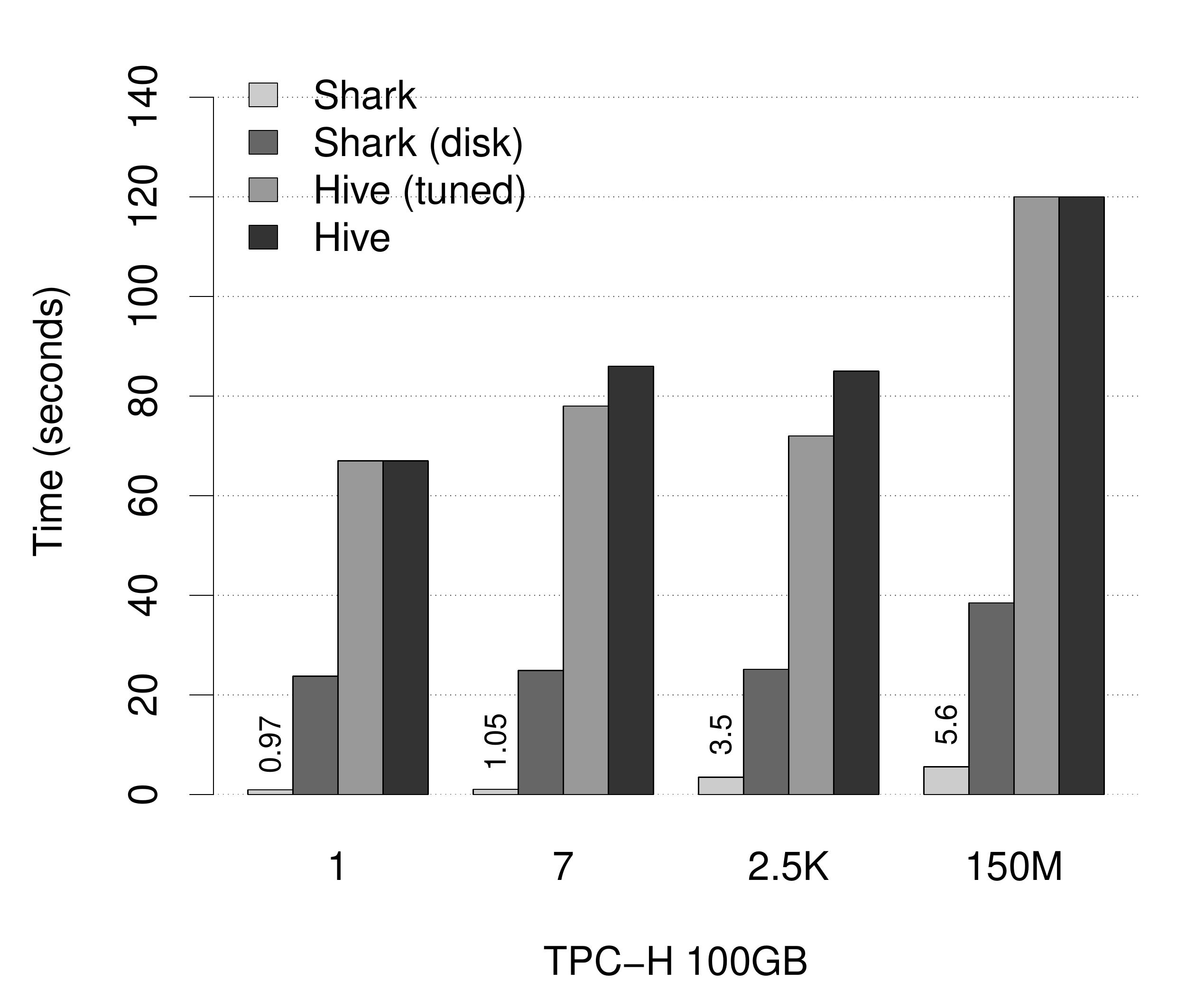}
  \end{minipage}
  \hspace{0.5cm}
  \begin{minipage}[b]{0.45\linewidth}
    \centering
    \includegraphics[width=\linewidth]{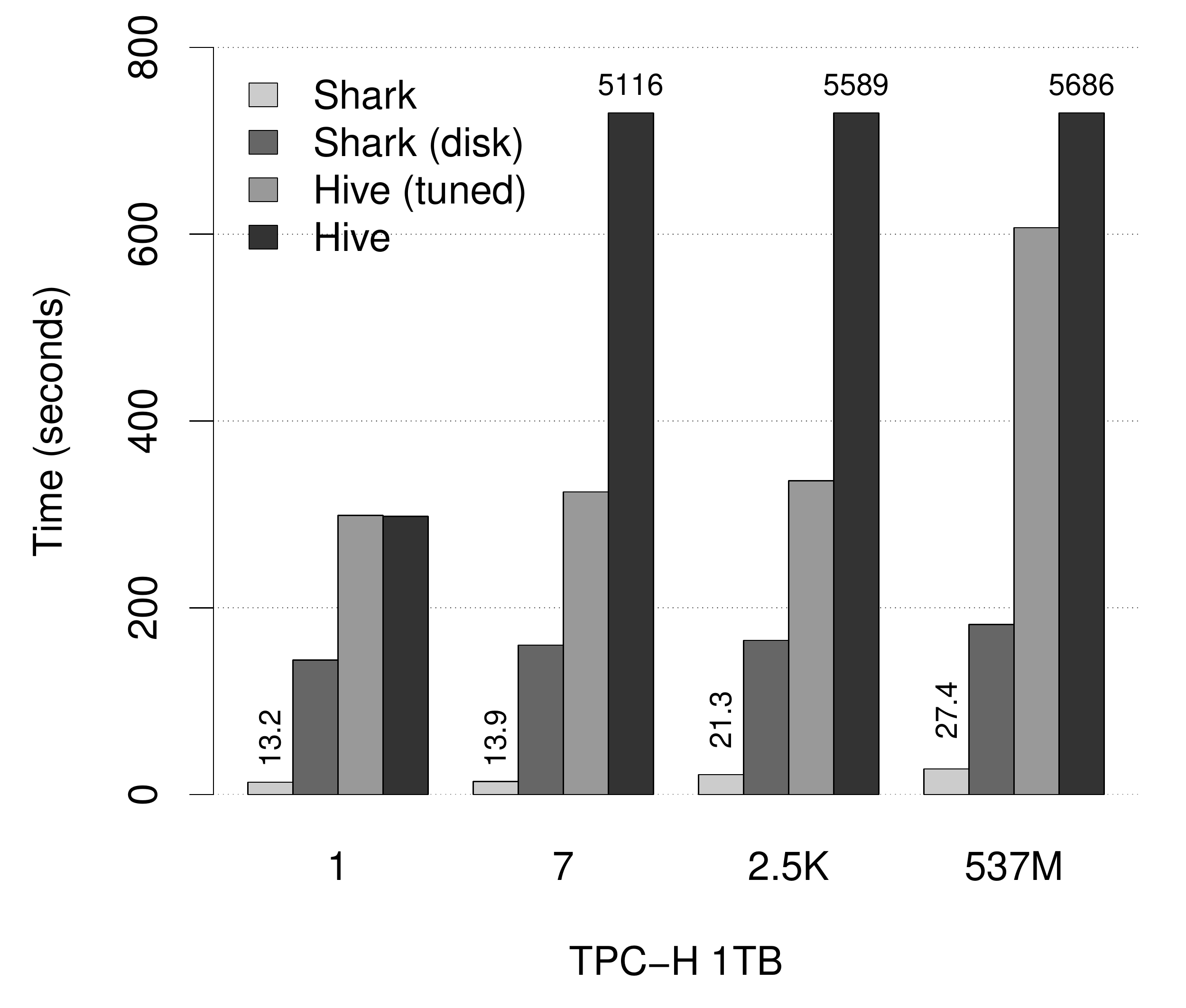}
  \end{minipage}
  \vspace{-10pt}
  \caption{Aggregation queries on lineitem table. X-axis indicates the number of groups for each aggregation query.}
  \label{fig:tpch-groupby}
\end{figure*}

The queries were of the form:
\lstset{basicstyle=\small\ttfamily}
\begin{lstlisting}
SELECT [GROUP_BY_COLUMN], COUNT(*) FROM lineitem
GROUP BY [GROUP_BY_COLUMN]
\end{lstlisting}

We chose to run one query with no group-by column (\ie a simple count), and three queries with group-by aggregations: SHIPMODE (7 groups), RECEIPTDATE (2500 groups), and SHIPMODE (150 million groups in 100 GB, and 537 million groups in 1 TB).

For both \system and Hive, aggregations were first performed on each partition, and then the intermediate aggregated results were partitioned and sent to reduce tasks to produce the final aggregation.
As the number of groups becomes larger, more data needs to be shuffled across the network.

Figure \ref{fig:tpch-groupby} compares the performance of \system and Hive, measuring \system's performance on both in-memory data and data loaded from HDFS.
As can be seen in the figure, \system was $80\times$ faster than hand-tuned Hive for queries with small numbers of groups, and $20\times$ faster for queries with large numbers of groups, where the shuffle phase domniated the total execution cost.

We were somewhat surprised by the performance gain observed for on-disk data in \system.
After all, both \system and Hive had to read data from HDFS and deserialize it for query processing.
This difference, however, can be explained by \system's very low task launching overhead, optimized shuffle operator, and other factors; see  Section~\ref{sec:discussion} for more details.

\subsubsection{Join Selection at Run-time}

\begin{figure}[t]
  \centering
  \includegraphics[width=\linewidth]{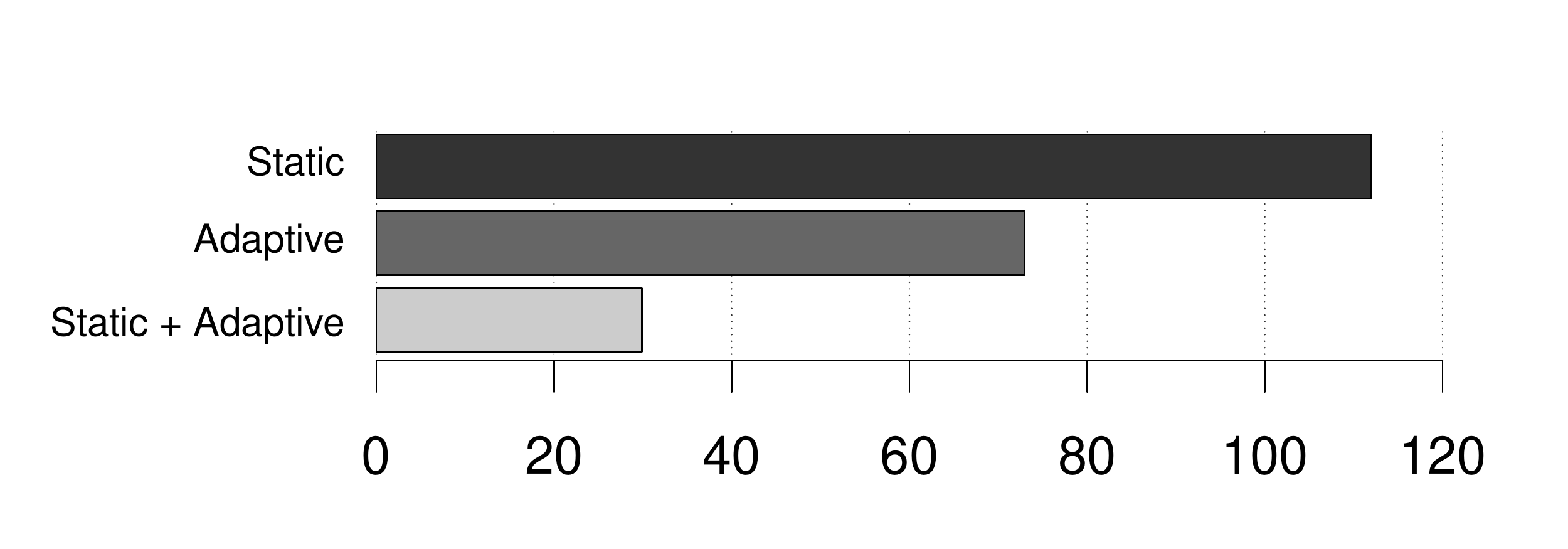}
  \vspace{-25pt}
  \caption{Join strategies chosen by optimizers (seconds)}
  \label{fig:join-strategies}
\end{figure}

In this experiment, we tested how \pde can improve query performance through run-time re-optimization of query plans.
The query joined the lineitem and supplier tables from the 1 TB TPC-H dataset,
using a UDF to select suppliers of interest based on their addresses.
In this specific instance, the UDF selected 1000 out of 10 million suppliers.
Figure~\ref{fig:join-strategies} summarizes these results.

\lstset{basicstyle=\small\ttfamily}
\begin{lstlisting}
SELECT * from lineitem l join supplier s
ON l.L_SUPPKEY = s.S_SUPPKEY
WHERE SOME_UDF(s.S_ADDRESS)
\end{lstlisting}

Lacking good selectivity estimation on the UDF, a static optimizer would choose to perform a shuffle join on these two tables because the initial sizes of both tables are large.
Leveraging \pde, after running the pre-shuffle map stages for both tables, \system's dynamic optimizer realized that the filtered supplier table was small.
It decided to perform a map-join, replicating the filtered supplier table to all nodes and performing the join using only map tasks on lineitem.

To further improve the execution, the optimizer can analyze the logical plan and infer that the probability of supplier table being small is much higher than that of lineitem (since supplier is smaller initially, and there is a filter predicate on supplier).
The optimizer chose to pre-shuffle only the supplier table, and avoided launching two waves of tasks on lineitem.
This combination of static query analysis and \pde led to a $3\times$ performance improvement over a na\"ive, statically chosen plan.

\subsubsection{Fault Tolerance}

\begin{figure}[t]
  \centering
  \includegraphics[width=\linewidth]{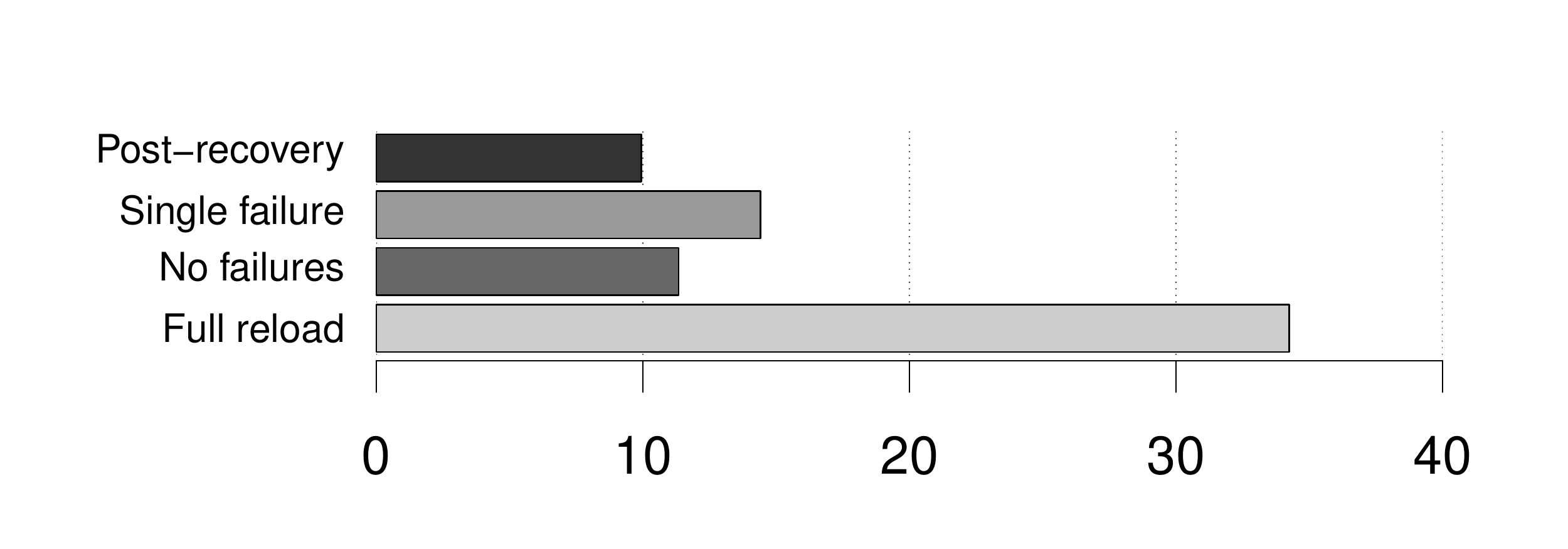}
  \vspace{-22pt}
  \caption{Query time with failures (seconds)}
  \label{fig:failure-recovery}
\end{figure}

To measure \system's performance in the presence of node failures, we simulated failures and measured query performance before, during, and after failure recovery.
Figure~\ref{fig:failure-recovery} summarizes fives runs of our failure recovery experiment, which was performed on a 50-node \texttt{m2.4xlarge} EC2 cluster.

We used a group-by query on the 100 GB lineitem table to measure query performance in the presence of faults.
After loading the lineitem data into \system's memory store, we killed a worker machine and re-ran the query.
\system gracefully recovered from this failure and parallelized the reconstruction of lost partitions on the other 49 nodes.
This recovery had a small performance impact ($\sim 3$ seconds), but it was significantly cheaper than the cost of re-loading the entire dataset and re-executing the query.

After this recovery, subsequent queries operated against the recovered dataset, albeit with fewer machines.
In Figure~\ref{fig:failure-recovery}, the post-recovery performance was marginally better than the pre-failure performance; we believe that this was a side-effect of the JVM's JIT compiler, as more of the scheduler's code might have become compiled by the time the post-recovery queries were run.

\subsection{Real Hive Warehouse Queries}

An early industrial user provided us with a sample of their Hive warehouse data and two years of query traces from their Hive system.
A leading video analytics company for content providers and publishers, the user built most of their analytics stack based on Hadoop.
The sample we obtained contained 30 days of video session data, occupying 1.7 TB of disk space when decompressed.
It consists of a single fact table containing 103 columns, with heavy use of complex data types such as \texttt{array} and \texttt{struct}.
The sampled query log contains 3833 analytical queries, sorted in order of frequency.
We filtered out queries that invoked proprietary UDFs and picked four frequent queries that are prototypical of other queries in the complete trace.
These queries compute aggregate video quality metrics over different audience segments:
\begin{packed_enum}
  \item Query 1 computes summary statistics in 12 dimensions for users of a specific customer on a specific day.
  \item Query 2 counts the number of sessions and distinct customer/client combination grouped by countries with filter predicates on eight columns.
  \item Query 3 counts the number of sessions and distinct users for all but 2 countries.
  \item Query 4 computes summary statistics in 7 dimensions grouping by a column, and showing the top groups sorted in descending order.
\end{packed_enum}

\begin{figure}[t]
  \centering
  \includegraphics[width=\linewidth]{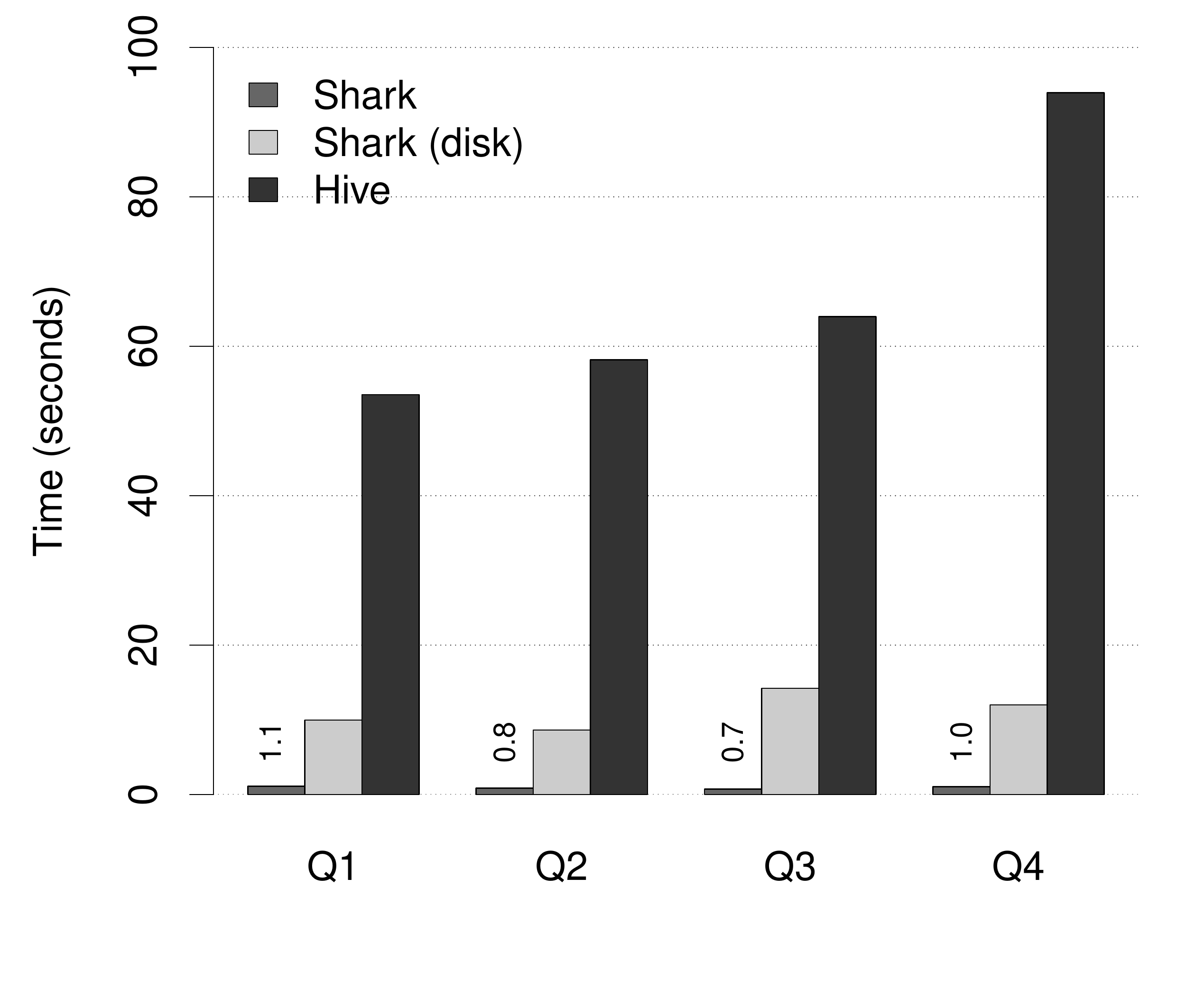}
  \vspace{-40pt}
  \caption{Real Hive warehouse workloads}
  \label{fig:videoanalytics}
\end{figure}

Figure \ref{fig:videoanalytics} compares the performance of \system and Hive on these queries.
The result is very promising as \system was able to process these real life queries in sub-second latency in all but one cases, whereas it took Hive 50 to 100 times longer to execute them.

A closer look into these queries suggests that this data exhibits the natural clustering properties mentioned in Section \ref{sec:mappruning}.
The map pruning technique, on average, reduced the amount of data scanned by a factor of 30.


\subsection{Machine Learning}
A key motivator of using SQL in a MapReduce environment is the ability to perform sophisticated machine learning on big data.
We implemented two iterative machine learning algorithms, logistic regression and k-means, to compare the performance of \system versus running the same workflow in Hive and Hadoop.

The dataset was synthetically generated and contained 1 billion rows and 10 columns, occupying 100 GB of space.
Thus, the feature matrix contained 1 billion points, each with 10 dimensions.
These machine learning experiments were performed on a 100-node \texttt{m1.xlarge} EC2 cluster.

Data was initially stored in relational form in \system's memory store and HDFS.
The workflow consisted of three steps: (1) selecting the data of interest from the warehouse using SQL, (2) extracting features, and (3) applying iterative algorithms.
In step 3, both algorithms were run for 10 iterations.

\begin{figure}[t]
  \centering
  \includegraphics[width=\linewidth]{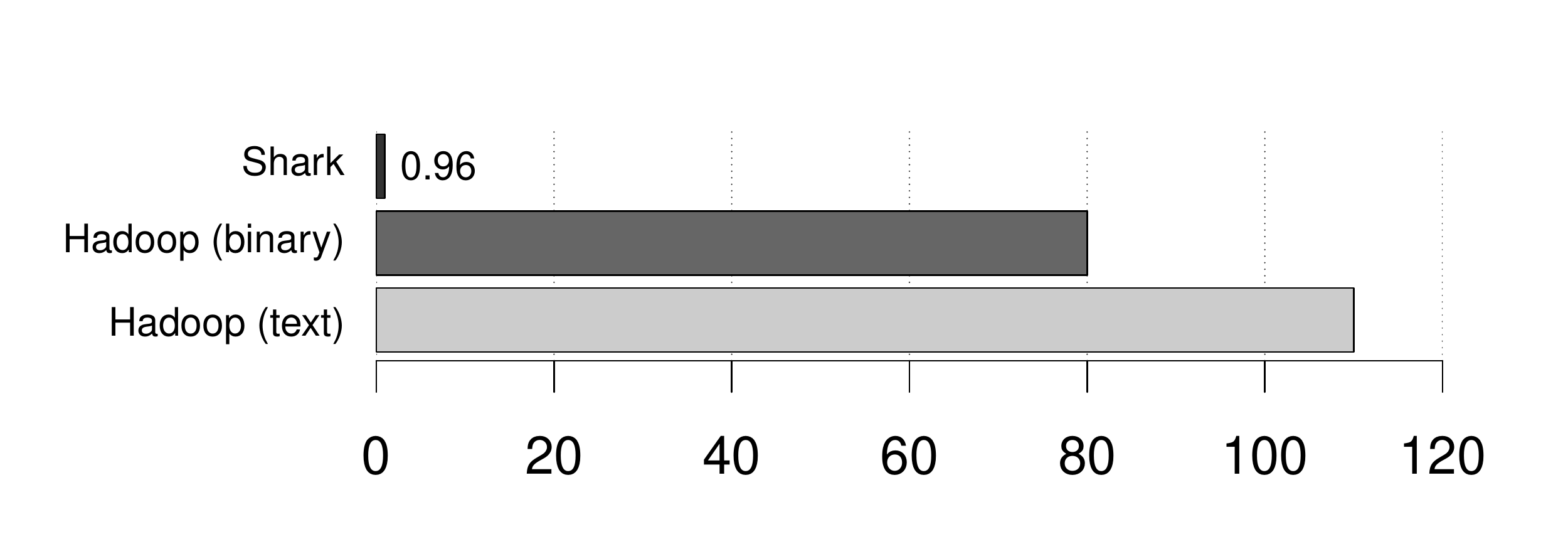}
  \vspace{-25pt}
  \caption{Logistic regression, per-iteration runtime (seconds)}
  \label{fig:logistic-regression}
  \vspace{-10pt}
\end{figure}

\begin{figure}[t]
  \centering
  \includegraphics[width=\linewidth]{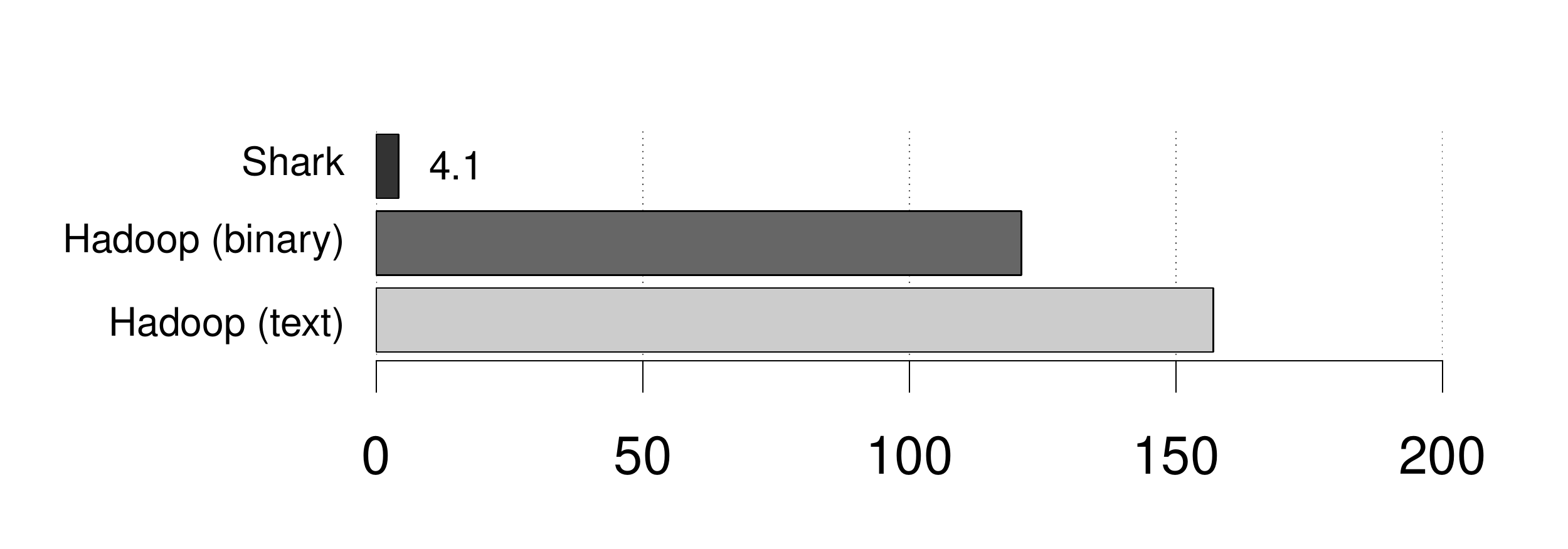}
  \vspace{-25pt}
  \caption{K-means clustering, per-iteration runtime (seconds)}
  \label{fig:k-means}
  \vspace{-10pt}
\end{figure}

Figures~\ref{fig:logistic-regression} and \ref{fig:k-means} show the time to execute a single iteration of logistic regression and k-means, respectively.
We implemented two versions of the algorithms for Hadoop, one storing input data as text in HDFS and the other using a serialized binary format.
The binary representation was more compact and had lower CPU cost in record deserialization, leading to improved performance.
Our results show that \system is $100\times$ faster than Hive and Hadoop for logistic regression and $30\times$ faster for k-means.
K-means experienced less speedup because it was computationally more expensive than logistic regression, thus making the workflow more CPU-bound.

In the case of \system, if data initially resided in its memory store, step 1 and 2 were executed in roughly the same time it took to run one iteration of the machine learning algorithm.
If data was not loaded into the memory store, the first iteration took 40 seconds for both algorithms.
Subsequent iterations, however, reported numbers consistent with Figures~\ref{fig:logistic-regression} and \ref{fig:k-means}.
In the case of Hive and Hadoop, every iteration took the reported time because data was loaded from HDFS for every iteration.

%% file: discussion.tex

\section{Discussion}
\label{sec:discussion}

\system shows that it is possible to run fast relational queries in a fault-tolerant manner
using the fine-grained deterministic task model introduced by MapReduce.
This design offers an effective way to scale query processing to ever-larger workloads, and
to combine it with rich analytics.
In this section, we consider two questions: first, why were previous MapReduce-based
systems, such as Hive, slow, and what gave \system its advantages?
Second, are there other benefits to the fine-grained task model? We argue that fine-grained
tasks also help with multitenancy and elasticity, as has been demonstrated in MapReduce
systems.

\subsection{Why are Previous MapReduce-Based Systems Slow?}

Conventional wisdom is that MapReduce is slower than MPP databases
for several reasons: expensive data materialization for fault tolerance, inferior data layout
(\eg lack of indices), and costlier execution strategies~\cite{pavlo2009comparison, mapreduce-dbms-cacm}.
Our exploration of Hive confirms these reasons, but also shows that a combination of conceptually simple
``engineering'' changes to the engine (\eg in-memory storage) and more involved architectural changes
(\eg \pde) can alleviate them.
We also find that a somewhat surprising variable not considered in detail in MapReduce systems,
the task scheduling overhead, actually has a dramatic effect on performance, and greatly improves
load balancing if minimized.

\vspace{4pt}\noindent\textbf{Intermediate Outputs:}
MapReduce-based query engines, such as Hive, materialize intermediate data to disk in two situations.
First, \emph{within} a MapReduce job, the map tasks save their output in case a reduce task
fails~\cite{mapreduce}. Second, many queries need to be compiled into \emph{multiple} MapReduce steps,
and engines rely on replicated file systems, such as HDFS, to store the output of each step.

For the first case, we note that map outputs were stored on disk primarily as a convenience to
ensure there is sufficient space to hold them in large batch jobs. Map outputs are \emph{not} replicated
across nodes, so they will still be lost if the mapper node fails~\cite{mapreduce}. Thus, if the outputs
fit in memory, it makes sense to store them in memory initially, and only spill them to disk if they are
large. \system's shuffle implementation does this by default, and sees far faster shuffle performance
(and no seeks) when the outputs fit in RAM. This is often the case in aggregations and
filtering queries that return a much smaller output than their input.\footnote{
Systems like Hadoop also benefit from the OS buffer cache in serving map outputs, but we found that
the extra system calls and file system journalling from writing map outputs to files still adds
overhead (Section~\ref{sec:memory-shuffle}).
}
Another hardware trend that may improve performance, even for large shuffles, is SSDs, which would
allow fast random access to a larger space than memory.

For the second case, engines that extend the MapReduce execution model to general task DAGs
can run multi-stage jobs without materializing any outputs to HDFS. Many such engines
have been proposed, including Dryad, Tenzing and Spark~\cite{dryad,tenzing,spark}.

\vspace{4pt}\noindent\textbf{Data Format and Layout:}
While the na\"ive pure schema-on-read approach to MapReduce incurs considerable processing
costs, many systems use more efficient storage formats within the MapReduce model to speed
up queries. Hive itself supports ``table partitions'' (a basic index-like system
where it knows that certain key ranges are contained in certain files, so it can avoid
scanning a whole table), as well as column-oriented representation of on-disk data~\cite{hive}.
We go further in \system by using fast in-memory columnar representations within Spark.
\system does this without modifying the Spark runtime by simply representing a block
of tuples as a single Spark record (one Java object from Spark's perspective), and
choosing its own representation for the tuples within this object.

Another feature of Spark that helps \system, but was not present in previous MapReduce
runtimes, is control over the data partitioning across
nodes (Section~\ref{sec:controlled-partitioning}). This lets us co-partition tables.

Finally, one capability of RDDs that we do not yet exploit is random reads. While RDDs only
support coarse-grained operations for their \emph{writes}, \emph{read} operations on them can be
fine-grained, accessing just one record~\cite{spark}. This would allow RDDs to be used as indices.
Tenzing can use such remote-lookup reads for joins~\cite{tenzing}.

\vspace{4pt}\noindent\textbf{Execution Strategies:}
Hive spends considerable time on sorting the data before each shuffle and writing the
outputs of each MapReduce stage to HDFS, both limitations of the rigid, one-pass MapReduce model
in Hadoop. More general runtime engines, such as Spark, alleviate some of these problems.
For instance, Spark supports hash-based distributed aggregation and general task DAGs.

To truly optimize the execution of relational queries, however, we found it necessary to
select execution plans based on data statistics. This becomes difficult in the presence
of UDFs and complex analytics functions, which we seek to support as first-class citizens
in \system. To address this problem, we proposed \pde (\PDEac), which allows
our modified version of Spark to \emph{change} the downstream portion of an execution graph
once each stage completes based on data statistics. \PDEac goes beyond the runtime graph rewriting
features in previous systems, such as DryadLINQ~\cite{dryadlinq}, by collecting fine-grained statistics
about ranges of keys and by allowing switches to a completely different join strategy, such
as broadcast join, instead of just selecting the number of reduce tasks.


\vspace{4pt}\noindent\textbf{Task Scheduling Cost:}
Perhaps the most surprising engine property that affected \system, however, was a purely
``engineering'' concern: the overhead of launching tasks.
Traditional MapReduce systems, such as Hadoop, were designed for multi-hour batch jobs
consisting of tasks that were several minutes long. They launched each task in a separate
OS process, and in some cases had a high latency to even submit a task.
For instance, Hadoop uses periodic ``heartbeats'' from each worker every 3 seconds
to assign tasks, and sees overall task startup delays of 5--10 seconds.
This was sufficient for batch workloads, but clearly falls short for ad-hoc queries.

Spark avoids this problem by using a fast event-driven RPC library to launch tasks and by reusing
its worker processes. It can launch thousands of tasks per second with only about 5 ms of overhead
per task, making task lengths of 50-100 ms and MapReduce jobs of 500 ms viable.
What surprised us is how much this affected query performance, even in large (multi-minute) queries.

Sub-second tasks allow the engine to balance work across nodes extremely well, even when some nodes
incur unpredictable delays (\eg network delays or JVM garbage collection).
They also help dramatically with skew. Consider, for example, a system that needs to run a hash
aggregation on 100 cores. If the system launches 100 reduce tasks, the key range for each task
needs to be carefully chosen, as any imbalance will slow down the entire job.
If it could split the work among 1000 tasks, then the slowest task can be as much as
$10\times$ slower than the average without affecting the job response time much!
After implementing skew-aware partition selection in \PDEac, we were somewhat disappointed that
it did not help compared to just having a higher number of reduce tasks in most workloads, because
Spark could comfortably support thousands of such tasks. However, this property makes the engine
highly robust to unexpected skew.

\begin{figure}[ht]
  \centering
  \begin{minipage}[b]{0.49\linewidth}
    \centering
    \includegraphics[width=\linewidth]{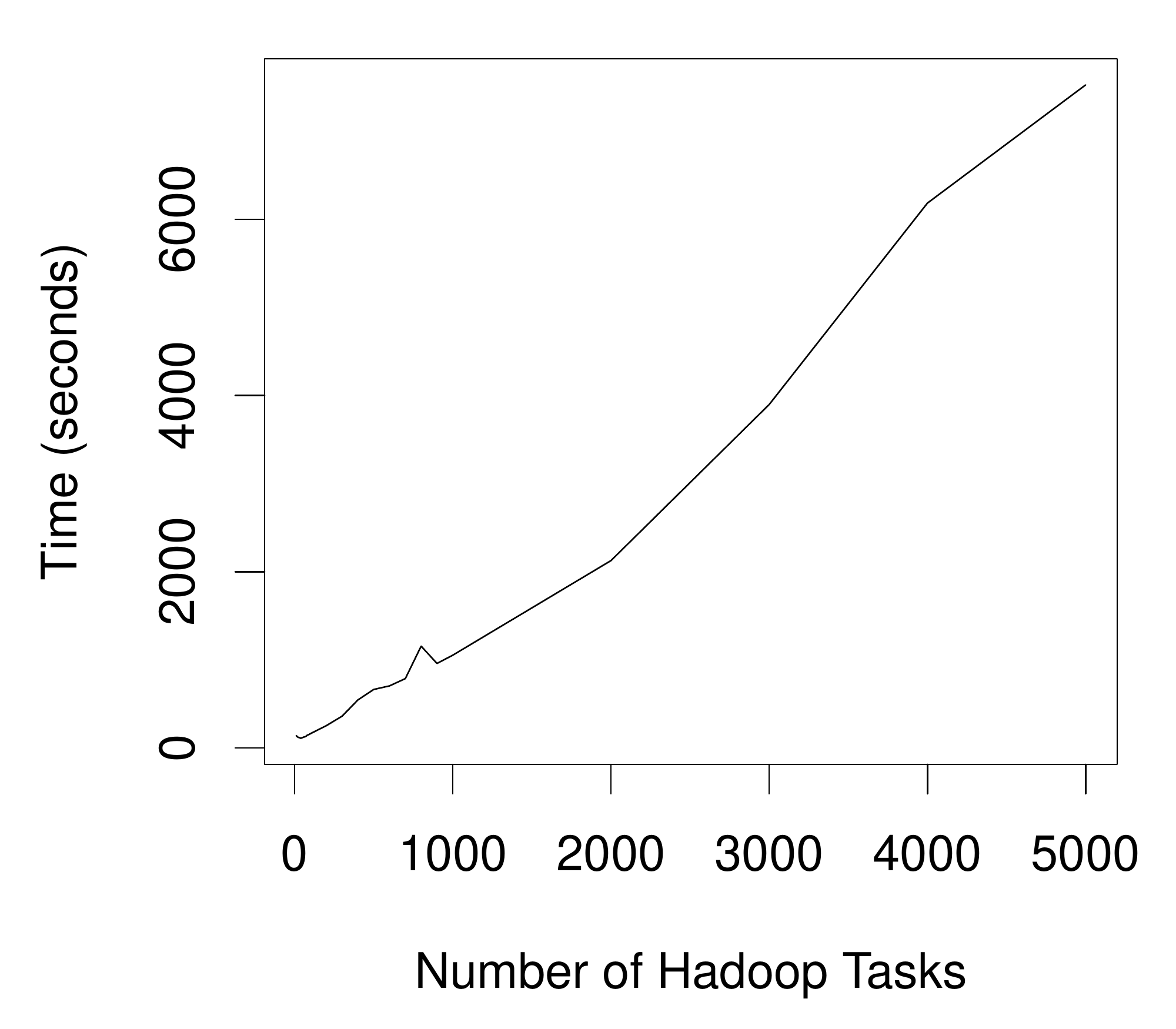}
  \end{minipage}
  \begin{minipage}[b]{0.49\linewidth}
    \centering
    \includegraphics[width=\linewidth]{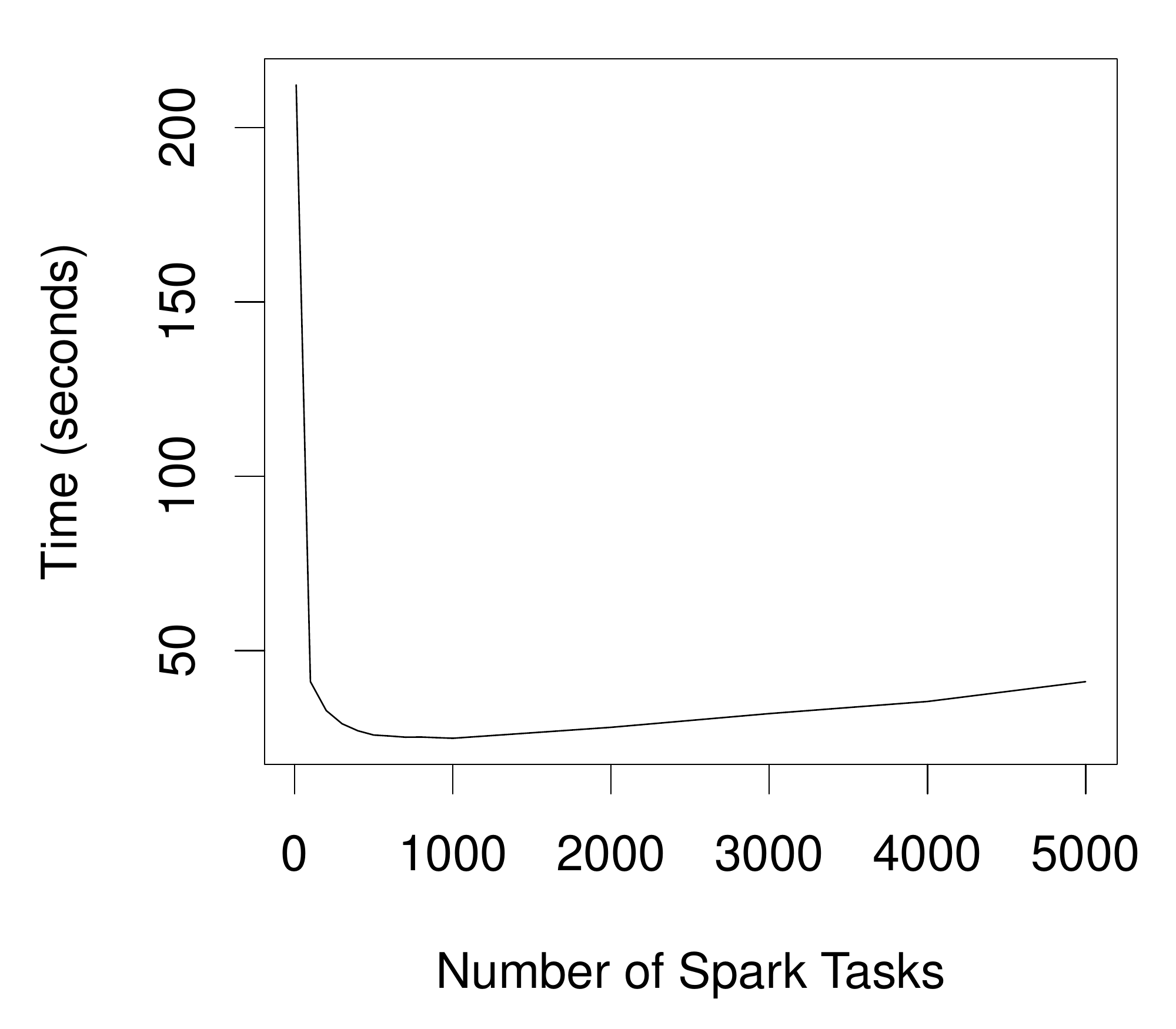}
  \end{minipage}
  \vspace{-10pt}
  \caption{Task launching overhead}
  \label{fig:task-overhead}
  \vspace{-10pt}
\end{figure}

In this way, Spark stands in contrast to Hadoop/Hive, where using the wrong number of tasks was
sometimes $10\times$ slower than an optimal plan, and there has been considerable work to
automatically choose the number of reduce tasks~\cite{skewtune, closer}.
Figure~\ref{fig:task-overhead} shows how job execution times varies as the number of reduce tasks launched by Hadoop and Spark. Since a Spark job can launch thousands of reduce tasks without incurring much overhead, partition data skew can be mitigated by always launching many tasks.

More fundamentally, there are few reasons why sub-second tasks should not be feasible
even at higher scales than we have explored, such as tens of thousands of nodes.
Systems like Dremel~\cite{dremel} routinely run sub-second, multi-thousand-node jobs.
Indeed, even if a single master cannot keep up with the scheduling decisions, the scheduling could be
delegated across ``lieutenant'' masters for subsets of the cluster.
Fine-grained tasks also offer many advantages over coarser-grained execution graphs beyond load
balancing, such as faster recovery (by spreading out lost tasks across more nodes) and query
elasticity; we discuss some of these next.

\subsection{Other Benefits of the Fine-Grained Task Model}

While this paper has focused primarily on the fault tolerance benefits of fine-grained deterministic
tasks, the model also provides other attractive properties. We wish to point out two benefits that
have been explored in MapReduce-based systems.

\vspace{4pt}\noindent\textbf{Elasticity:}
In traditional MPP databases, a distributed query plan is selected once, and the system needs to run
at that level of parallelism for the whole duration of the query.
In a fine-grained task system, however, nodes can appear or go away during a query, and pending work
will automatically be spread onto them.
This enables the database engine to naturally be elastic. If an administrator wishes to remove nodes
from the engine (\eg in a virtualized corporate data center), the engine can simply treat those as
failed, or (better yet) proactively replicate their data to other nodes if given a few minutes' warning.
Similarly, a database engine running on a cloud could scale \emph{up} by requesting new VMs if a query is
expensive. Amazon's Elastic MapReduce~\cite{amazon-emr} already supports resizing clusters at runtime.

\vspace{4pt}\noindent\textbf{Multitenancy:}
The same elasticity, mentioned above, enables dynamic resource sharing between users. In a traditional
MPP database, if an important query arrives while another large query using most of the cluster, there are few options beyond canceling the earlier query. In systems based on fine-grained
tasks, one can simply wait a few seconds for the current tasks from the first query to finish, and
start giving the nodes tasks from the second query.
For instance, Facebook and Microsoft have developed
fair schedulers for Hadoop and Dryad that allow large historical queries, compute-intensive machine
learning jobs, and short ad-hoc queries to safely coexist~\cite{delay-scheduling,quincy}.


%% file: related_work.tex

\section{Related Work}
\label{sec:related}

To the best of our knowledge, \system is the only low-latency system that can efficiently combine SQL and machine learning workloads, while supporting fine-grained fault recovery.

We categorize large-scale data analytics systems into three classes. First, systems like Hive~\cite{hive}, Tenzing~\cite{tenzing}, SCOPE~\cite{scope}, and Cheetah~\cite{cheetah} compile declarative queries into MapReduce-style jobs. Even though some of them introduce modifications to the execution engine they are built on, it is hard for these systems to achieve interactive query response times for reasons discussed in Section \ref{sec:discussion}.

Second, several projects aim to provide low-latency engines using architectures resembling shared-nothing parallel databases. Such projects include PowerDrill~\cite{powerdrill} and Impala~\cite{impala}. These systems do not support fine-grained fault tolerance.
In case of mid-query faults, the entire query needs to be re-executed.
Google's Dremel~\cite{dremel} does rerun lost tasks, but it only supports an aggregation tree topology for query execution,
and not the more complex shuffle DAGs required for large joins or distributed machine learning.

A third class of systems take a hybrid approach by combining a MapReduce-like engine with relational databases. HadoopDB~\cite{hadoopdb} connects multiple single-node database systems using Hadoop as the communication layer. Queries can be parallelized using Hadoop MapReduce, but within each MapReduce task, data processing is pushed into the relational database system. Osprey~\cite{osprey} is a middleware layer that adds fault-tolerance properties to parallel databases.
It does so by breaking a SQL query into multiple small queries and sending them to parallel databases for execution. \system presents a much simpler single-system architecture that supports all of the properties of this third class of systems, as well as statistical learning capabilities that HadoopDB and Osprey lack.

The \pde (\PDEac) technique introduced by \system resembles adaptive query optimization techniques
proposed in \cite{eddies, query-scrambling, dewitt-mid-query}. It is, however, unclear how these single-node
techniques would work in a distributed setting and scale out to hundreds of nodes.
In fact, \PDEac actually complements some of these techniques, as \system can use \PDEac to optimize how data gets shuffled \emph{across} nodes, and use the traditional single-node techniques \emph{within} a local task.
DryadLINQ~\cite{dryadlinq} optimizes its number of reduce tasks at run-time based on map output sizes, but does not collect richer statistics, such as histograms, or make broader execution plan changes, such as changing join algorithms, like \PDEac can.
RoPE~\cite{rope} proposes using historical query information to optimize query plans, but relies on repeatedly executed queries. \PDEac works on queries that are executing for the first time.

Finally, \system builds on the distributed approaches for machine learning developed in systems like
Graphlab~\cite{graphlab}, Haloop~\cite{haloop}, and Spark~\cite{spark}. However, \system is unique in
offering these capabilities in a SQL engine, allowing users to select data of interest using SQL
and immediately run learning algorithms on it without time-consuming export to another system.
Compared to Spark, \system also provides far more efficient in-memory representation of relational
data, and mid-query optimization using \PDEac.

%% file: conclusion.tex

\section{Conclusion}
\label{sec:conclusion}

We have presented \system, a new data warehouse system that combines fast relational queries
and complex analytics in a single, fault-tolerant runtime.
\system generalizes a MapReduce-like runtime to run SQL effectively, using both traditional
database techniques, such as column-oriented storage, and a novel \emph{\pde (\PDEac)}
technique that lets it reoptimize queries at run-time based on fine-grained data statistics.
This designs enables \system to generally match the speedups reported for MPP databases over
MapReduce, while simultaneously providing machine learning functions in the same engine
and fine-grained, mid-query fault tolerance across both SQL and machine learning.
Overall, the system is up to $100\times$ faster than Hive for SQL, and $100\times$
faster than Hadoop for machine learning.

We have open sourced \system at \url{shark.cs.berkeley.edu}, and have also worked with two Internet
companies as early users. They report speedups of 40--100$\times$ on real
queries, consistent with our results.

%% file: paper.bbl
\begin{thebibliography}{10}

\bibitem{impala}
https://github.com/cloudera/impala.

\bibitem{amazon-emr}
http://aws.amazon.com/about-aws/whats-new/2010/10/20/amazon-elastic-mapreduce-introduces-resizing-running-job-flows/.

\bibitem{hadoopdb}
A.~Abouzeid et~al.
\newblock Hadoopdb: an architectural hybrid of mapreduce and dbms technologies
  for analytical workloads.
\newblock {\em VLDB}, 2009.

\bibitem{rope}
S.~Agarwal et~al.
\newblock Re-optimizing data-parallel computing.
\newblock In {\em NSDI'12}.

\bibitem{pacman}
G.~Ananthanarayanan et~al.
\newblock Pacman: Coordinated memory caching for parallel jobs.
\newblock In {\em NSDI}, 2012.

\bibitem{eddies}
R.~Avnur and J.~M. Hellerstein.
\newblock Eddies: continuously adaptive query processing.
\newblock In {\em SIGMOD}, 2000.

\bibitem{haloop}
Y.~Bu et~al.
\newblock {HaLoop}: efficient iterative data processing on large clusters.
\newblock {\em Proc. VLDB Endow.}, 2010.

\bibitem{scope}
R.~Chaiken et~al.
\newblock Scope: easy and efficient parallel processing of massive data sets.
\newblock {\em VLDB}, 2008.

\bibitem{tenzing}
B.~Chattopadhyay, , et~al.
\newblock Tenzing a sql implementation on the mapreduce framework.
\newblock {\em PVLDB}, 4(12):1318--1327, 2011.

\bibitem{cheetah}
S.~Chen.
\newblock Cheetah: a high performance, custom data warehouse on top of
  mapreduce.
\newblock {\em VLDB}, 2010.

\bibitem{mapreduce-ml}
C.~Chu et~al.
\newblock Map-reduce for machine learning on multicore.
\newblock {\em Advances in neural information processing systems}, 19:281,
  2007.

\bibitem{madlib}
J.~Cohen, B.~Dolan, M.~Dunlap, J.~Hellerstein, and C.~Welton.
\newblock Mad skills: new analysis practices for big data.
\newblock {\em VLDB}, 2009.

\bibitem{mapreduce}
J.~Dean and S.~Ghemawat.
\newblock {{MapReduce}}: Simplified data processing on large clusters.
\newblock In {\em OSDI}, 2004.

\bibitem{unified-db-analytics}
X.~Feng et~al.
\newblock Towards a unified architecture for in-rdbms analytics.
\newblock In {\em SIGMOD}, 2012.

\bibitem{closer}
B.~Guffler et~al.
\newblock Handling data skew in mapreduce.
\newblock In {\em CLOSER}, 2011.

\bibitem{powerdrill}
A.~Hall et~al.
\newblock Processing a trillion cells per mouse click.
\newblock {\em VLDB}.

\bibitem{dryad}
M.~Isard et~al.
\newblock Dryad: distributed data-parallel programs from sequential building
  blocks.
\newblock {\em SIGOPS}, 2007.

\bibitem{quincy}
M.~Isard et~al.
\newblock Quincy: Fair scheduling for distributed computing clusters.
\newblock In {\em SOSP '09}, 2009.

\bibitem{dryadlinq}
M.~Isard and Y.~Yu.
\newblock Distributed data-parallel computing using a high-level programming
  language.
\newblock In {\em SIGMOD}, 2009.

\bibitem{dewitt-mid-query}
N.~Kabra and D.~J. DeWitt.
\newblock Efficient mid-query re-optimization of sub-optimal query execution
  plans.
\newblock In {\em SIGMOD}, 1998.

\bibitem{skewtune}
Y.~Kwon et~al.
\newblock Skewtune: mitigating skew in mapreduce applications.
\newblock In {\em SIGMOD '12}, 2012.

\bibitem{graphlab}
Y.~Low et~al.
\newblock Distributed graphlab: a framework for machine learning and data
  mining in the cloud.
\newblock {\em VLDB}, 2012.

\bibitem{pregel}
G.~Malewicz et~al.
\newblock Pregel: a system for large-scale graph processing.
\newblock In {\em SIGMOD}, 2010.

\bibitem{dremel}
S.~Melnik et~al.
\newblock Dremel: interactive analysis of web-scale datasets.
\newblock {\em Proc. VLDB Endow.}, 3:330--339, Sept 2010.

\bibitem{pavlo2009comparison}
A.~Pavlo et~al.
\newblock A comparison of approaches to large-scale data analysis.
\newblock In {\em SIGMOD}, 2009.

\bibitem{mapreduce-dbms-cacm}
M.~Stonebraker et~al.
\newblock Mapreduce and parallel dbmss: friends or foes?
\newblock {\em Commun. ACM}.

\bibitem{cstore}
M.~Stonebraker et~al.
\newblock C-store: a column-oriented dbms.
\newblock In {\em VLDB}, 2005.

\bibitem{hive}
A.~Thusoo et~al.
\newblock Hive-a petabyte scale data warehouse using hadoop.
\newblock In {\em ICDE}, 2010.

\bibitem{tpch}
Transaction Processing Performance Council.
\newblock {\em TPC BENCHMARK H}.

\bibitem{query-scrambling}
T.~Urhan, M.~J. Franklin, and L.~Amsaleg.
\newblock Cost-based query scrambling for initial delays.
\newblock In {\em SIGMOD}, 1998.

\bibitem{osprey}
C.~Yang et~al.
\newblock Osprey: Implementing mapreduce-style fault tolerance in a
  shared-nothing distributed database.
\newblock In {\em ICDE}, 2010.

\bibitem{delay-scheduling}
M.~Zaharia et~al.
\newblock Delay scheduling: A simple technique for achieving locality and
  fairness in cluster scheduling.
\newblock In {\em {EuroSys} 10}, 2010.

\bibitem{spark}
M.~Zaharia et~al.
\newblock Resilient distributed datasets: a fault-tolerant abstraction for
  in-memory cluster computing.
\newblock NSDI, 2012.

\end{thebibliography}
